%% file: Main_preprint.tex
\UseRawInputEncoding
\documentclass[twocolumn,showpacs,preprintnumbers,amsmath,amssymb]{revtex4-1}

\usepackage{graphicx}
\usepackage{dcolumn}
\usepackage[T1]{fontenc}
\usepackage{bm}
\usepackage{amsmath}
\usepackage{color}
\usepackage{soul}
\usepackage{gensymb}
\usepackage[colorlinks=true, citecolor=blue, linkcolor=blue, urlcolor=blue]{hyperref}
\newcommand{\correct}[2]{{\color{red}\sout{#1}\,}{\color{blue}#2}}
\begin{document}

\title{Unified Weak-to-Strong Coupling Transitions and Radiation Interference Induced by Vertical-Symmetry Breaking in Photonic Crystal Slabs}

\author{H. H. Chu$^{1,2}$}
\author{R. Mermet-Lyaudoz$^3$}
\author{F. Dubois$^3$}
\author{S. H. Nguyen$^4$}
\author{Q. M. Bui$^5$}
\author{E. Drouard$^3$}
\author{L. Berguiga$^3$}
\author{C. Seassal$^3$}
\author{X. Letartre$^3$}
\author{P. Viktorovitch$^3$}
\author{C. Dang$^{1,2}$}
\author{Q. Le-Van$^{4,5,6,7}$}
\author{H. S. Nguyen$^{3,2,8}$}
\email{hai-son.nguyen@ec-lyon.fr}

\affiliation{$^1$School of Electrical and Electronic Engineering, Nanyang Technological University, Singapore 639798, Singapore}
\affiliation{$^2$CNRS-International-NTU-Thales Research Alliance (CINTRA), IRL 3288, Singapore 637553, Singapore}
\affiliation{$^3$Universit\'e de Lyon, Institut des Nanotechnologies de Lyon, INL/CNRS, \'Ecole Centrale de Lyon, 36 avenue Guy de Collongue, 69130 Ecully, France}
\affiliation{$^4$Center for Environmental Intelligence, VinUniversity, Hanoi 100000, Vietnam}
\affiliation{$^5$VinUni-Illinois Smart Health Center, VinUniversity, Hanoi 100000, Vietnam}
\affiliation{$^6$Center for Materials Innovation and Technology, VinUniversity, Hanoi 100000, Vietnam}
\affiliation{$^7$College of Engineering and Computer Sciences, VinUniversity, Hanoi 100000, Vietnam}
\affiliation{$^8$Institut Universitaire de France (IUF), 75231 Paris, France}

\date{\today}

\begin{abstract}
Vertical-symmetry breaking provides a versatile means of coupling leaky photonic-crystal resonances that are otherwise protected by opposite out-of-plane parity. Here, we develop a unified two-mode framework for the weak-to-strong coupling transition induced by vertical-symmetry breaking in photonic crystal slabs. The symmetry-breaking perturbation simultaneously generates a near-field coherent coupling and modifies the overlap of the radiation channels of the two parent resonances. Their interplay drives the transition from frequency crossings to avoided crossings through an exceptional point (EP), while also governing radiative linewidth exchange through Friedrich-Wintgen interference. Using a radiation-vector description resolved into the upper and lower half-spaces, we show that total radiation cancellation and one-sided radiation cancellation correspond, respectively, to quasi-bound states in the continuum (quasi-BIC) and quasi-unidirectional guided resonances (quasi-UGR). This framework distinguishes the spectral condition for EP formation from the far-field conditions controlling linewidth suppression and directional emission. We validate this picture numerically and experimentally in square-lattice photonic crystal slabs. Tuning the superstrate index drives a weak-to-strong coupling transition through an EP, while partial etching yields a broad off-$\Gamma$ quasi-BIC regime with strongly asymmetric top and bottom radiation. We further apply the same radiation-vector framework to a laterally shifted bilayer grating, where quasi-BICs and quasi-UGR emerge along a continuous symmetry-breaking pathway. These results establish vertical-symmetry breaking as a general route for controlling hybridization, radiation interference, and directional leakage in photonic crystal slabs.
\end{abstract}

\maketitle

\section{Introduction}
Leaky resonances in photonic crystal (PhC) slabs offer a powerful setting for controlling dispersion, radiation, and non-Hermitian mode coupling above the light line~\cite{Fan2002,Johnson1999,Fan2003}. They can host ultra-high-$Q$ states, structured far fields, and tunable radiative interactions, enabling surface-emitting sources, nonlinear optics, sensing, directional emission, and integrated optical interfaces. Among the most prominent manifestations are bound states in the continuum
(BICs) ~\cite{Friedrich1985,Hsu2013,Plotnik2011,Zhen2014,Hsu2016,Koshelev2018,Trinh2022,Nguyen2024_vin}, exceptional points (EPs)~\cite{Heiss2012,Zhen2015} and unidirectional guided resonances (UGRs) ~\cite{Ji2026,Yin2023,Yuan2026,Yin2020,Zeng2021}, which enable ultra-high-$Q$ resonances~\cite{Huang2023},
structured radiation singularities~\cite{Yuan2025,Sang2022}, enhanced sensing~\cite{Chen2017,Wiersig2014}, and unconventional lasing
dynamics~\cite{Liertzer2012,Brandstetter2014}. These phenomena are often discussed separately: EPs are identified through the topology of complex eigenfrequencies, Friedrich-Wintgen (FW) BICs through destructive interference in the total radiated field, and UGRs through cancellation of radiation into one selected half-space. 

\begin{figure}[ht!]
    \centering
    \includegraphics[width=\linewidth]{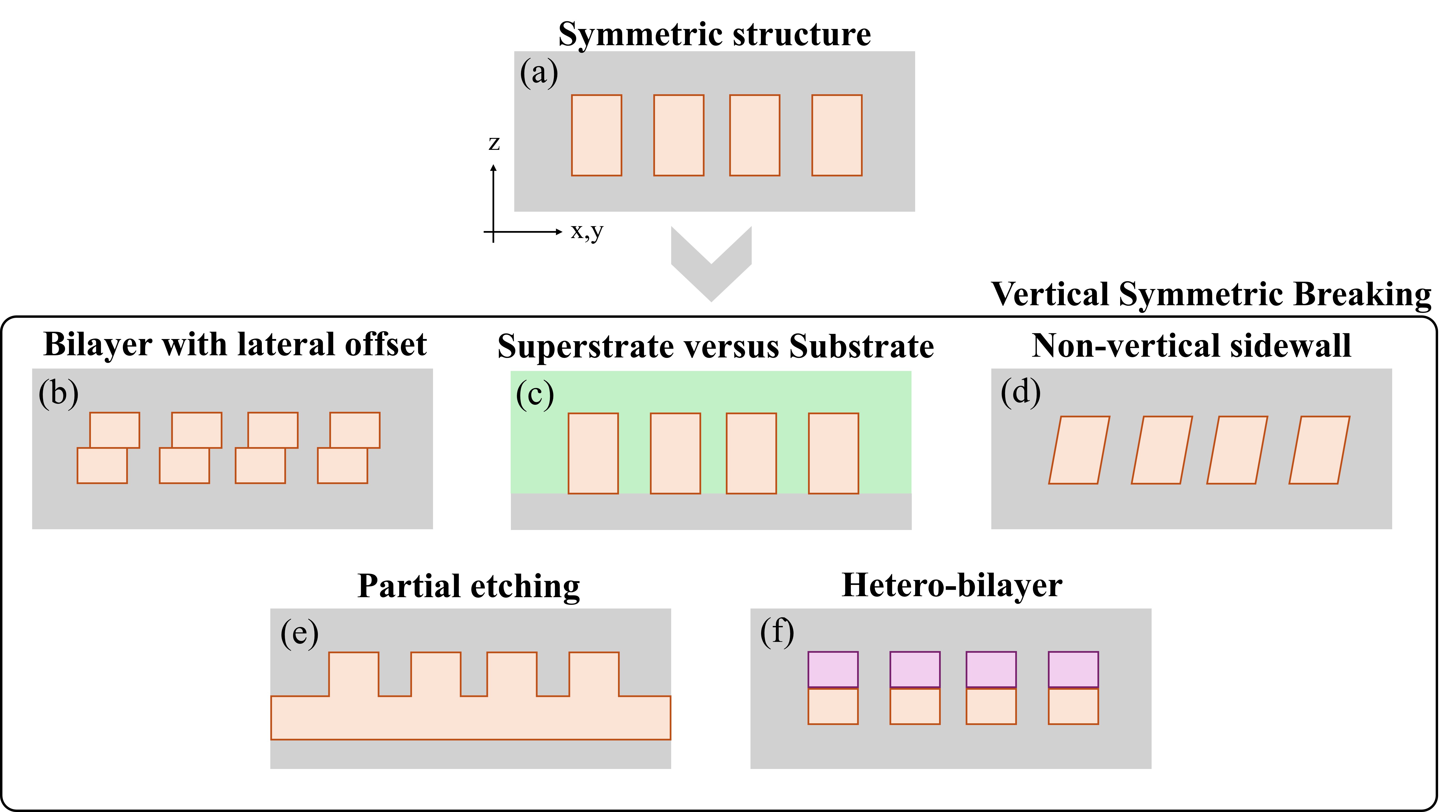}
  \caption{Representative routes to vertical-symmetry breaking in photonic crystal slabs.
(a) Vertically symmetric reference structure, where the superstrate and substrate are identical and the structure preserves the \(z\!\to\!-z\) mirror symmetry.
(b--f) Five common scenarios that break vertical symmetry:
(b) a bilayer of identical patterned slabs with a lateral offset;
(c) asymmetric superstrate/substrate refractive indices;
(d) non-vertical sidewalls;
(e) partial etching leaving a residual unpatterned layer;
and (f) an aligned hetero-bilayer composed of dissimilar slabs with different materials or thicknesses.
The coupled-mode framework developed in this work applies to all of these configurations, as it relies only on symmetry breaking and radiation-channel coupling.
}
    \label{fig:sketch}
\end{figure}

A particularly general route for controlling leaky resonances is the breaking of vertical, or out-of-plane, mirror symmetry. In a vertically symmetric PhC slabs, resonances can be classified according to their parity under reflection across the mid-plane~\cite{Hsu2017}. Modes of opposite vertical parity radiate into orthogonal combinations of the top and bottom channels, which strongly constrain their non-Hermitian coupling~\cite{Fan2003,WonjooSuh2004}. Breaking this vertical symmetry relaxes these constraints and generically induces coupling between modes that would otherwise remain decoupled~\cite{Yin2023,Nguyen2018,Wang2018}. Vertical-symmetry breaking can arise in several experimentally relevant ways. Starting from the vertically symmetric reference structure shown in Figure~\ref{fig:sketch}(a)~\cite{Letartre2022}, vertical symmetry can be broken through several experimentally relevant routes. In multilayer platforms, it can be introduced by laterally shifting two identical patterned layers [Figure~\ref{fig:sketch}(b)]~\cite{Wang2018,Ni2024,Choi2025,Nguyen2026,Zeng2021,Lee2022_plr,Saadi2025,Duc2026}. In single-layer PhC slabs, vertical symmetry can be broken by placing the slab on a substrate whose refractive index differs from that of the superstrate [Figure~\ref{fig:sketch}(c)]~\cite{Gowadzka2021,Tian2020}, by using non-vertical sidewalls [Figure~\ref{fig:sketch}(d)]~\cite{Yin2020,Yin2023,YuanLu2025,Li2025}, or by partially etching the photonic pattern so that a residual unpatterned layer remains [Figure~\ref{fig:sketch}(e)]~\cite{Nguyen2018,Cueff2019}. Finally, in multilayer heterostructures, vertical symmetry can also be broken by forming an aligned hetero-bilayer composed of layers with different material properties or thicknesses [Figure~\ref{fig:sketch}(f)]~\cite{Park2024,Yuan2026}. Although these configurations appear geometrically distinct, they share a common physical mechanism: vertical-symmetry breaking modifies the coherent modal coupling, the total radiative interference, and the side-resolved radiation into the upper and lower half-spaces.

Previous studies have shown, in specific vertically asymmetric geometries, that opposite-parity mode hybridization can generate avoided crossings, off-$\Gamma$ quasi-BICs, or directional guided resonances~\cite{Nguyen2018,Yin2023,Lee2024,Lee2025,Yuan2026}. However, the weak-to-strong coupling transition, FW linewidth exchange, and direction-resolved radiation cancellation are commonly treated as separate effects, often using geometry-specific models. In particular, the weak-to-strong coupling transition and EP formation are characterized through the complex eigenvalue spectrum, whereas quasi-BICs and UGRs are identified through total or side-resolved radiation properties. A common framework is needed to clarify which features are governed by spectral hybridization, which arise from total-radiation interference, and which require the full direction-resolved radiation information.

Here, we develop a unified two-mode non-Hermitian description of this generic vertical-symmetry-breaking mechanism. Starting from two leaky resonances of opposite vertical parity, we show that vertical-symmetry breaking simultaneously generates near-field coherent coupling and radiative-channel mixing. The resulting complex spectrum undergoes a weak-to-strong coupling transition through an EP. The same hybridization enables FW redistribution of radiative loss, which can strongly suppress the total radiative decay and produce a quasi-BIC. Resolving the hybrid radiation vector into upper and lower components further distinguishes total-radiation cancellation from one-sided cancellation: the former yields an ideal BIC or, more generally, a quasi-BIC, whereas the latter yields a UGR or quasi-UGR. In this picture, EPs, quasi-BICs, and UGRs are not identical phenomena, but distinct spectral or radiation conditions imposed on the same two-mode hybrid system.

We first derive the general model and identify the effective quantities governing the weak-to-strong coupling transition and the attainable FW linewidth redistribution. We then validate the framework in a square-lattice SiN$_x$-on-SiO$_2$ PhC slab. Full-wave simulations show that tuning the superstrate refractive index drives the system continuously from weak to strong coupling through an EP, while partial etching reshapes the radiative interference and produces a broad off-$\Gamma$ quasi-BIC regime. Angle-resolved reflectivity measurements reveal the associated symmetry-breaking quasi-BIC near the hybridization region and are reproduced by the same two-mode model. Finally, we apply the same radiation-vector description to a laterally shifted bilayer grating, where vertical and in-plane symmetry breaking cooperate to connect a symmetry-protected BIC, quasi-UGRs, and quasi-BICs through a continuous radiation-interference pathway. 

\section{General Theoretical Framework: From Symmetry Breaking to Non-Hermitian Coupling}\label{sec:General_theory}

We consider a photonic crystal slab periodic in the in-plane directions, where electromagnetic eigenmodes are labeled by the in-plane Bloch wave vector $k_{\parallel}$ in the first Brillouin zone. Fixing $k_{\parallel}$, we focus on two leaky Bloch resonances above the light line with mode profiles $\mathbf{E}_m(\mathbf r)$ and complex eigenfrequencies $\tilde{\omega}_m=\omega_m-i\gamma_m$, with $m=1,2$. We adopt the time-dependence convention $e^{-i\omega t}$, under which a positive loss rate appears as a negative imaginary contribution to the eigenfrequency. Thus, $\gamma_m$ is the amplitude decay rate, corresponding to a spectral half-width at half-maximum, while the full linewidth is $2\gamma_m$ and the quality factor is $Q_m=\omega_m/(2\gamma_m)$.

To establish the physical origin of the coupling, we first introduce a vertically symmetric reference structure with dielectric profile $\varepsilon_0(x,y,z)=\varepsilon_0(x,y,-z)$. In this symmetric limit, the resonances possess definite vertical parity. Consequently, even resonances radiate symmetrically toward the top and bottom half-spaces, whereas odd resonances radiate antisymmetrically. The two resonances therefore couple to orthogonal radiation subspaces ~\cite{WonjooSuh2004}. This radiation orthogonality enforces the cancellation of the dissipative coupling, while parity also forbids coherent coupling between the two modes. The effective Hamiltonian of the symmetric reference system is therefore diagonal:
\begin{equation}
H_0=
\begin{pmatrix}
\omega_1^{(0)}-i\gamma_1^{(0)} & 0 \\
0 & \omega_2^{(0)}-i\gamma_2^{(0)}
\end{pmatrix}.
\label{eq:H0}
\end{equation}

We now introduce a perturbation $\Delta\varepsilon(\mathbf r)$ that breaks the vertical mirror symmetry. This perturbation simultaneously induces a coherent, or dispersive, coupling $\kappa$ and modifies the radiation amplitudes into the external channels. As a result, the radiation vectors of the two modes are no longer orthogonal, and a finite dissipative coupling is generated through radiative-channel mixing. Absorbing diagonal frequency shifts and linewidth changes into $\omega_{1,2}$ and $\gamma_{1,2}$, the effective non-Hermitian Hamiltonian in the basis of the two unperturbed resonances takes the form
\begin{equation}
H_{\mathrm{eff}}=
\begin{pmatrix}
\omega_1-i\gamma_1 & \kappa-i\Gamma \\
\kappa-i\Gamma^* & \omega_2-i\gamma_2
\end{pmatrix},
\qquad
\kappa\in\mathbb R^+ .
\label{eq:Heff-vsb}
\end{equation}
Here $\kappa$ is the coherent coupling induced by vertical-symmetry breaking, while $\Gamma$ is the off-diagonal dissipative coupling associated with the non-orthogonality of the outgoing radiation channels. We choose the relative phase of the two modal basis states such that the coherent coupling $\kappa$ is real and positive. With this gauge choice, the physically relevant relative phase between coherent and dissipative coupling is carried by $\Gamma$. The quantities $\omega_{1,2}$ and $\gamma_{1,2}$ denote the diagonal frequency shifts and radiative decay rates of the perturbed parent resonances, while the observable hybrid resonances are obtained by diagonalizing $H_{\mathrm{eff}}$.

The effective Hamiltonian in Eq.\eqref{eq:Heff-vsb} captures the hybridization of two vertically symmetric parent resonances once the mirror symmetry is broken. A Hamiltonian of this form was used in Ref.\cite{Le2024}; here, we go beyond this phenomenological description by deriving a microscopic interpretation of each term and by relating the coherent and dissipative couplings to the perturbation $\Delta\varepsilon(\mathbf r)$. In particular, we show that the coherent coupling $\kappa$ originates from a near-field overlap between the two unperturbed modes, whereas the dissipative coupling $\Gamma$ arises from the far-field overlap of their radiation vectors. To this end, we first introduce for each resonance a radiation vector $\mathbf K_m$ whose components are the outgoing amplitudes into all open radiation channels: $\mathbf K_m = \left(K_{m,\alpha}\right)_\alpha$, with $\alpha=(s,\mathbf G,\sigma)$. Here $s=t,b$ denotes emission into the top or bottom half-space, $\mathbf G$ labels an open diffraction order, and $\sigma$ labels the transverse polarization of the outgoing plane wave. With the convention $\tilde{\omega}_m=\omega_m-i\gamma_m$, the radiative part of the effective Hamiltonian is written as $-(i/2)K^\dagger K$. Therefore,
\begin{equation}
\gamma_m = \frac{1}{2}\mathbf K_m^\dagger\mathbf K_m
= \frac{1}{2}\sum_{\alpha}|K_{m,\alpha}|^2 .
\label{eq:gamma-radiation-vector}
\end{equation}

As mentioned previously, the diagonal quantities $\omega_m$ and $\gamma_m$ are modified from $\omega_m^{(0)}$ and $\gamma_m^{(0)}$ by the perturbation. To first order in $\Delta\varepsilon$, the diagonal frequency shift of mode $m$ is
\begin{equation}
\begin{gathered}
\delta\omega_m \simeq -\frac{\omega_m^{(0)}}{2} \frac{\int_V \Delta\varepsilon(\mathbf r) |\mathbf E_m^{(0)}(\mathbf r)|^2\,dV}{\int_V \varepsilon_0(\mathbf r) |\mathbf E_m^{(0)}(\mathbf r)|^2\,dV}, \\
\omega_m = \omega_m^{(0)} + \delta\omega_m .
\end{gathered}
\label{eq:diagonal-frequency-shift}
\end{equation}
Here $\mathbf E_1^{(0)}$ and $\mathbf E_2^{(0)}$ are the uncoupled even-like and odd-like modes of the vertically symmetric reference structure.

The diagonal radiative decay rate is obtained from the perturbed radiation vector $\mathbf K_m=\mathbf K_m^{(0)}+\delta\mathbf K_m$ as $\gamma_m \simeq \gamma_m^{(0)} + \mathrm{Re}\left[(\mathbf K_m^{(0)})^\dagger\delta\mathbf K_m
\right]$, where $(\mathbf K_m^{(0)})^\dagger\delta\mathbf K_m=\sum_\alpha K_{m,\alpha}^{(0)*}\delta K_{m,\alpha}$ and the second-order term $\delta\mathbf K_m^\dagger\delta\mathbf K_m/2$ has been omitted. The radiation-amplitude correction can be expressed schematically as
\begin{equation}
\delta K_{m,\alpha} \propto \int_V
\Delta\varepsilon(\mathbf r)\,
\mathbf E_m^{(0)}(\mathbf r)\cdot
\mathbf E_\alpha^{\rm rad}(\mathbf r)\,dV ,
\label{eq:radiation-amplitude-perturbation}
\end{equation}
where $\mathbf E_\alpha^{\rm rad}$ is the outgoing radiation channel associated with $\alpha=(s,\mathbf G,\sigma)$. Thus, $\Delta\varepsilon$ modifies both the resonance frequencies and the radiative linewidths of the parent modes before the final hybridization is obtained by diagonalizing the Hamiltonian~\eqref{eq:Heff-vsb}.

Crucially, the same perturbation $\Delta\varepsilon(\mathbf r)$ also generates the off-diagonal coherent coupling. Microscopically, this coupling can be estimated from spatial coupled-mode perturbation theory ~\cite{Huang1994,Tanaka2003} as
\begin{equation}
\kappa = \frac{\omega_0^{(0)}}{2} \frac{\left|\int_V \mathbf E_1^{(0)*}(\mathbf r)\cdot
\mathbf E_2^{(0)}(\mathbf r)\, \Delta\varepsilon(\mathbf r)\,dV \right| }{
\sqrt{\int_V|\mathbf E_1^{(0)}(\mathbf r)|^2\varepsilon_0(\mathbf r)\,dV}
\sqrt{\int_V|\mathbf E_2^{(0)}(\mathbf r)|^2\varepsilon_0(\mathbf r)\,dV}
}.
\label{eq:kappa_overlap}
\end{equation}
Here $\omega_0^{(0)}=\frac{\omega_1^{(0)}+\omega_2^{(0)}}{2}$. Equation~\eqref{eq:kappa_overlap} shows that $\kappa$ vanishes by parity in the vertically symmetric limit and becomes finite once $\Delta\varepsilon(\mathbf r)$ breaks the $z\rightarrow -z$ mirror symmetry. The sign of $\kappa$ is gauge-dependent and has been fixed by the convention used in Eq.~\eqref{eq:Heff-vsb}.

The off-diagonal dissipative coupling is the corresponding cross-overlap of the two radiation vectors, given by $\Gamma=\frac{1}{2}\mathbf K_1^\dagger\mathbf K_2
= \frac{1}{2}\sum_{\alpha}K_{1,\alpha}^{*}K_{2,\alpha}$. Introducing the normalized radiation vectors $\hat{\mathbf K}_m=\mathbf K_m/\sqrt{2\gamma_m}$, we obtain
\begin{equation}
\Gamma =  \sqrt{\gamma_1\gamma_2}\,\rho e^{i\phi}, 
\label{eq:Gamma12-rho-phi}
\end{equation}
with $\rho e^{i\phi} = \hat{\mathbf K}_1^\dagger\hat{\mathbf K}_2$. The Cauchy--Schwarz inequality gives $0\leq\rho\leq1$. Thus, $\rho$ measures the degree of collinearity between the outgoing radiation patterns of the two resonances, represented by the radiation vectors $\mathbf K_1$ and $\mathbf K_2$ in the multi-channel radiation space, while $\phi$ is the global phase of their radiative overlap. In the gauge of Eq.~\eqref{eq:Heff-vsb}, $\phi$ is the relative phase between the dissipative coupling $\Gamma$ and the real coherent coupling $\kappa$.

Microscopically, it is useful to decompose the radiation vectors into their top and bottom components, $\mathbf K_m=(\mathbf K_m^t,\mathbf K_m^b)^T$, where each component still contains all open diffraction orders and polarizations. The dissipative coupling then becomes
$\Gamma=\frac{1}{2}(\mathbf K_1^t)^\dagger\mathbf K_2^t+\frac{1}{2}
(\mathbf K_1^b)^\dagger\mathbf K_2^b$. Defining the direction-resolved decay rates $\gamma_m^{t,b}=(\mathbf K_m^{t,b})^\dagger\mathbf K_m^{t,b}/2$, with $\gamma_m=\gamma_m^t+\gamma_m^b$, where $\gamma_m^{t,b}$ denotes the total radiative decay rate of mode $m$ into the top or bottom half-space after summing over all open diffraction orders and polarizations, and defining the direction-resolved overlaps
$\rho_{t,b}e^{i\phi_{t,b}}=(\mathbf K_1^{t,b})^\dagger\mathbf K_2^{t,b}
/\left(2\sqrt{\gamma_1^{t,b}\gamma_2^{t,b}}\right),$ one obtains
\begin{equation}
\Gamma=\rho_t\sqrt{\gamma_1^t\gamma_2^t}e^{i\phi_t} +\rho_b\sqrt{\gamma_1^b\gamma_2^b}e^{i\phi_b}.
\label{eq:rho-top-bottom}
\end{equation}
This expression shows explicitly that the effective dissipative coupling results from the coherent sum of the top and bottom radiative overlaps. In the vertically symmetric limit with opposite vertical parity, the top and bottom radiation vectors are related by $\mathbf K_1^t=\mathbf K_1^b$ for the even-like mode and $\mathbf K_2^t=-\mathbf K_2^b$ for the odd-like mode. Hence, $(\mathbf K_1^t)^\dagger\mathbf K_2^t + (\mathbf K_1^b)^\dagger\mathbf K_2^b =0$, so the off-diagonal radiative self-energy $\Gamma$ vanishes. Vertical-symmetry breaking removes this cancellation by modifying both the relative amplitudes $\gamma_m^{t,b}$ and the relative phases $\phi_{t,b}$ of the top and bottom radiation channels, thereby generating a finite $\Gamma$.

In the sub-diffractive regime, where only the zeroth diffraction order is open, the radiation vector is simplified but does not reduce to a single scalar channel: it still contains the top and bottom outgoing waves. If, in addition, the relevant resonances couple predominantly to a single transverse polarization, for example the zeroth-order $s$-polarized channel, then each direction-resolved radiation vector becomes a scalar amplitude, $K_m^t$ or $K_m^b$. In that case the direction-resolved overlaps have unit modulus, $\rho_t=\rho_b=1$, provided both modes radiate into the corresponding half-space. However, the global overlap $\rho$ is not necessarily equal to unity, because the top and bottom contributions can still interfere constructively or destructively. This leads to 
\begin{equation}
\Gamma_{(\text{single channel})}=\sqrt{\gamma_1^t\gamma_2^t}e^{i\phi_t}+\sqrt{\gamma_1^b\gamma_2^b}e^{i\phi_b}.    \end{equation}
 Thus, in the single-polarization zeroth-order limit, polarization and diffraction-channel mismatch are negligible, and the nontrivial dissipative coupling is governed by the top/bottom radiation imbalance and by the relative phase between the top and bottom radiative overlaps.

The non-Hermitian Hamiltonian~\eqref{eq:Heff-vsb} therefore captures the central physical mechanism of this work: vertical-symmetry breaking simultaneously induces coherent hybridization, quantified by the near-field perturbative overlap in Eq.~\eqref{eq:kappa_overlap}, and radiative-channel mixing, quantified by the far-field overlap in Eq.~\eqref{eq:Gamma12-rho-phi}. Depending on the relative strength of the complex off-diagonal coupling $\kappa-i\Gamma$ and the losses contrast $\gamma_1-\gamma_2$, the two resonances can undergo weak or strong coupling. The boundary between these regimes corresponds to an EP, where the two complex eigenfrequencies and eigenvectors coalesce, as demonstrated in the following section. 

\begin{table*}[ht!]
\caption{Summary of the main symbols of the two-mode non-Hermitian model.}
\label{tab:notation}
\centering
\begin{tabular}{ll}
\hline
Symbol & Meaning \\
\hline
$\omega_{1,2}$, $\gamma_{1,2}$ & Frequencies and (amplitude) decay rates of the two bare modes \\
$\omega_0=(\omega_1+\omega_2)/2$ & Mean frequency \\
$\gamma_0=(\gamma_1+\gamma_2)/2$ & Mean decay rate \\
$\Delta=(\omega_1-\omega_2)/2$ & Half frequency detuning (swept via $k_\parallel$) \\
$\delta=(\gamma_1-\gamma_2)/2$ & Half linewidth (loss) contrast \\
$\kappa\in\mathbb{R}^+$ & Coherent (near-field) coupling, Eq.~\eqref{eq:kappa_overlap} \\
$\Gamma=\sqrt{\gamma_1\gamma_2}\,\rho e^{i\phi}$ & Dissipative (radiative) coupling, Eq.~\eqref{eq:Gamma12-rho-phi} \\
$\rho\in[0,1]$ & Modulus of the normalized radiation-vector overlap \\
$\phi$ & Relative phase of the radiative overlap \\
$P=(\kappa-i\Gamma)(\kappa-i\Gamma^*)$ & Off-diagonal coupling product, Eq.~\eqref{eq:P-product} \\
$\kappa_{\mathrm{eff}}=|\mathrm{Re}\sqrt{P}\,|$ & Effective coherent coupling, Eq.~\eqref{eq:keff-mu} \\
$\mu=(\mathrm{Im}\sqrt{P})^2/(\gamma_1\gamma_2)\in[0,1]$ & Radiation-overlap parameter
        ($\to\rho^2$ for real $\Gamma$, with $\mu\le\rho^2$), Eq.~\eqref{eq:keff-mu} \\
$\eta=\min(\gamma_1,\gamma_2)/\max(\gamma_1,\gamma_2)$ & Bare linewidth (loss) ratio \\
$\kappa_c$ & Critical coupling at the weak--strong boundary / EP, Eq.~\eqref{eq:weak-strong-criterion} \\
$\mathbf{K}_m$, $\mathbf{K}_m^{t,b}$ & Radiation vector of mode $m$ (top/bottom components) \\
$C$ & Normalized Friedrich--Wintgen loss-exchange contrast, Eq.~\eqref{eq:C-eta-phi} \\
$\mathcal{D}_{\mathbf c}$ & Direction-resolved (top/bottom) radiation contrast, Eq.~\eqref{eq:UGR-contrast} \\
\hline
\end{tabular}
\end{table*}

The linewidths of the observable hybrid resonances are obtained from the imaginary parts of the eigenvalues of $H_{\mathrm{eff}}$, or equivalently from the radiation vectors of the corresponding eigenmodes. For a hybrid eigenmode with coefficient vector $\mathbf c=(c_1,c_2)^T$, the corresponding radiation vector is $\mathbf K_{\mathbf c}=c_1\mathbf K_1+c_2\mathbf K_2$. Its total radiative decay rate is then $\gamma_{\mathbf c}=\mathbf K_{\mathbf c}^\dagger\mathbf K_{\mathbf c}/2$, 
with analogous expressions for the top and bottom contributions obtained from the corresponding top and bottom radiation components. Destructive interference between the radiation amplitudes of the two parent resonances can therefore suppress the total radiative decay, producing a BIC in the ideal case or a quasi-BIC when the cancellation is incomplete. The same direction-resolved formulation also provides a natural criterion for unidirectional guided resonances. A hybrid resonance is dark toward the top but remains radiative toward the bottom when the top contribution to $\gamma_{\mathbf c}$ vanishes while the bottom contribution remains finite, and conversely for bottom-dark emission. Thus, beyond EPs and quasi-BICs, vertical-symmetry breaking can also enable quasi-UGRs through destructive interference in only one radiation half-space. These different regimes -- weak and strong coupling, EPs, quasi-BIC formation, and quasi-UGRs -- together with their physical consequences and experimental manifestations, will also be discussed in detail in the following sections. For convenience, the symbols of the two-mode model used throughout this work are collected in Table~\ref{tab:notation}.

\subsection{From Weak to Strong Coupling: Frequency Crossings, EPs, and Avoided Crossings}
\label{sec:weak-strong-fw}

The complex eigenvalues of Eq.~\eqref{eq:Heff-vsb} determine both the resonance 
dispersions and their radiative linewidths. Introducing the offset frequency 
$\omega_0=(\omega_1+\omega_2)/2$, the energy detuning $\Delta=(\omega_1-\omega_2)/2$, 
the offset loss rate $\gamma_0=(\gamma_1+\gamma_2)/2$, and the loss contrast 
$\delta=(\gamma_1-\gamma_2)/2$, the complex eigenvalues can be written as
\begin{equation}
\tilde{\omega}_\pm(\Delta)
=
\omega_0-i\gamma_0
\pm
\alpha(\Delta),
\label{eq:lambda-pm}
\end{equation}
where the complex gap $\alpha(\Delta)$ is defined as
\begin{equation}
\alpha(\Delta)
=
\sqrt{
(\Delta-i\delta)^2
+
(\kappa-i\Gamma)(\kappa-i\Gamma^*)
}.
\label{eq:alpha}
\end{equation}
The real parts $\mathrm{Re}\,\tilde{\omega}_\pm$ give the hybrid-mode frequencies, while 
the linewidths are $\gamma_\pm=-\mathrm{Im}\,\tilde{\omega}_\pm$. In practice, the detuning 
$\Delta$ is controlled by sweeping the in-plane wavevector $k_{\parallel}$ near the 
bare band crossing, while $\kappa$, $\Gamma$, and $\gamma_{1,2}$ remain approximately 
constant over the relevant range.

The transition between weak and strong coupling can be defined operationally by whether the real parts of the eigenvalues cross for real detuning. A real-frequency crossing occurs if and only if there exists a real detuning $\Delta_\times$
such that $\Delta\omega(\Delta_\times)=0$, i.e.\ $\mathrm{Re}\alpha(\Delta_\times)=0$. Solving this condition gives the crossing position
\begin{equation}
\Delta_\times=
-\frac{\kappa\,\mathrm{Re}\Gamma}{\delta},
\label{eq:Delta-cross}
\end{equation}
provided that the coherent coupling remains below the critical value
\begin{equation}
\kappa_c=
|\delta|
\sqrt{
\frac{\delta^2+|\Gamma|^2}
{\delta^2+(\mathrm{Re}\Gamma)^2}
}.
\label{eq:weak-strong-criterion}
\end{equation}
We thus define:
\begin{itemize}
\item[(i)] \textbf{Weak-coupling regime} ($|\kappa|<\kappa_c$): there exists a real detuning $\Delta_\times$ given by Eq.~\eqref{eq:Delta-cross}
at which the hybrid-mode frequencies cross, $\Delta\omega(\Delta_\times)=0$.
\item[(ii)] \textbf{Strong-coupling regime} ($|\kappa|>\kappa_c$): $\mathrm{Re}\alpha(\Delta)>0$ for all real $\Delta$, so the hybrid-mode frequencies never
cross and a finite frequency gap persists throughout the detuning sweep.
\end{itemize}
As shown in Eq.~\eqref{eq:weak-strong-criterion}, the dissipative coupling affects the generalized threshold through $|\Gamma|$ and $\mathrm{Re}\Gamma$, and also controls {where} in detuning space the crossing occurs, when it exists, through Eq.~\eqref{eq:Delta-cross}.

\begin{figure*}[ht!]
    \centering
    \includegraphics[width=1\linewidth]{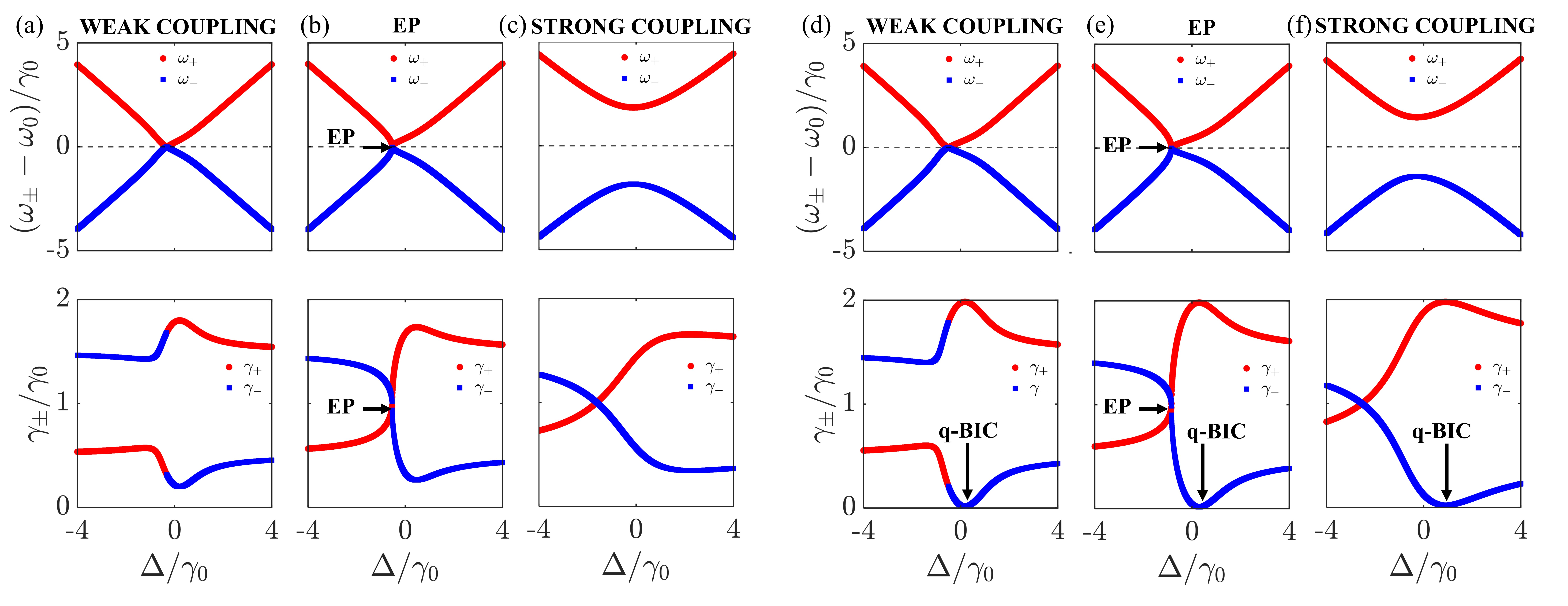}
   \caption{
\textbf{Weak-to-strong coupling transition and Friedrich--Wintgen quasi-BIC formation in the two-mode non-Hermitian model.}
(a--c) Real-frequency shifts \((\omega_\pm-\omega_0)/\gamma_0\) and normalized linewidths \(\gamma_\pm/\gamma_0\) as the detuning \(\Delta/\gamma_0\) is swept for moderate radiation-channel overlap. The system evolves from weak coupling in (a), where the real parts cross while the linewidths remain split, to the EP in (b), where both the frequencies and linewidths coalesce, and to strong coupling in (c), where the real parts anticross while the linewidths exchange between the two branches. 
(d--f) Same transition for near-collinear radiation vectors, where Friedrich--Wintgen interference produces a pronounced quasi-BIC linewidth minimum. In this high-overlap regime, the quasi-BIC can appear in the weak-coupling regime (d), at the EP (e), and in the strong-coupling regime (f), demonstrating that quasi-BIC formation is governed primarily by radiation-channel overlap rather than by the coherent avoided crossing gap. 
Parameters are \(\delta/\gamma_0=0.5\). For (a--c), \(\rho=0.8\) and \(\phi=0.3\pi\); for (d--f), \(\rho=0.98\) and \(\phi=0.05\pi\). The coherent coupling is chosen as \(\kappa=0.6\kappa_c\), \(\kappa=\kappa_c\), and \(\kappa=3\kappa_c\) for the weak-coupling, EP, and strong-coupling cases, respectively.
}
    \label{fig:transition_noQuasiBIC}
\end{figure*}

It is instructive to consider the limit of negligible dissipative coupling, $\Gamma\to0$ (orthogonal radiation channels). Equation~\eqref{eq:weak-strong-criterion} then reduces to $\kappa_c=|\delta|=|\gamma_1-\gamma_2|/2$, so that strong coupling sets in when $\kappa>|\delta|$. This recovers the textbook strong-coupling criterion that the coherent splitting must exceed the loss contrast of the bare resonances~\cite{Rodriguez2016}. A finite radiative overlap merely renormalizes this threshold through the factor $\sqrt{(\delta^2+|\Gamma|^2)/(\delta^2+(\mathrm{Re}\,\Gamma)^2)}$ appearing in Eq.~\eqref{eq:weak-strong-criterion}, so that our definition of the weak-to-strong coupling transition coincides with the conventional one in the purely coherent limit.

At the boundary between these two regimes, the two eigenvalues and eigenvectors coalesce ~\cite{Heiss2012,Miri2019}. This EP is obtained by imposing $\alpha(\Delta_{\mathrm{EP}})=0$, which yields
\begin{equation}
\kappa=\kappa_c,
\qquad
\Delta_{\mathrm{EP}}
=
-\frac{\kappa_c\,\mathrm{Re}\Gamma}{\delta}.
\label{eq:EP-condition}
\end{equation}
The EPs therefore lie exactly on the weak-to-strong coupling boundary. The consequences of Eqs.~\eqref{eq:Delta-cross}--\eqref{eq:EP-condition} are illustrated in Figures~\ref{fig:transition_noQuasiBIC}(a)--\ref{fig:transition_noQuasiBIC}(f). Figures~\ref{fig:transition_noQuasiBIC}(a) and \ref{fig:transition_noQuasiBIC}(d) show the weak-coupling regime, where $\kappa=0.6\kappa_c<\kappa_c$ and the real parts of the eigenvalues cross while the linewidths remain split. Figures~\ref{fig:transition_noQuasiBIC}(b) and \ref{fig:transition_noQuasiBIC}(e) corresponds to the transition between the two regimes, $\kappa=\kappa_c$, where an EP emerges and the real and imaginary parts of the eigenvalues coalesce simultaneously. Figures~\ref{fig:transition_noQuasiBIC}(c) and \ref{fig:transition_noQuasiBIC}(f) show the strong-coupling regime, where $\kappa=3\kappa_c>\kappa_c$ and the real parts anticross while the linewidths exchange between the two hybrid branches. The distinction between the two cases lies in the radiation-channel overlap: Figs.~\ref{fig:transition_noQuasiBIC}(a)--(c) use moderate overlap, whereas Figs.~\ref{fig:transition_noQuasiBIC}(d)--(f) use near-collinear radiation vectors, resulting in a much stronger radiative loss exchange between the hybrid branches.

\begin{figure}[ht!]
    \centering
    \includegraphics[width=1\linewidth]{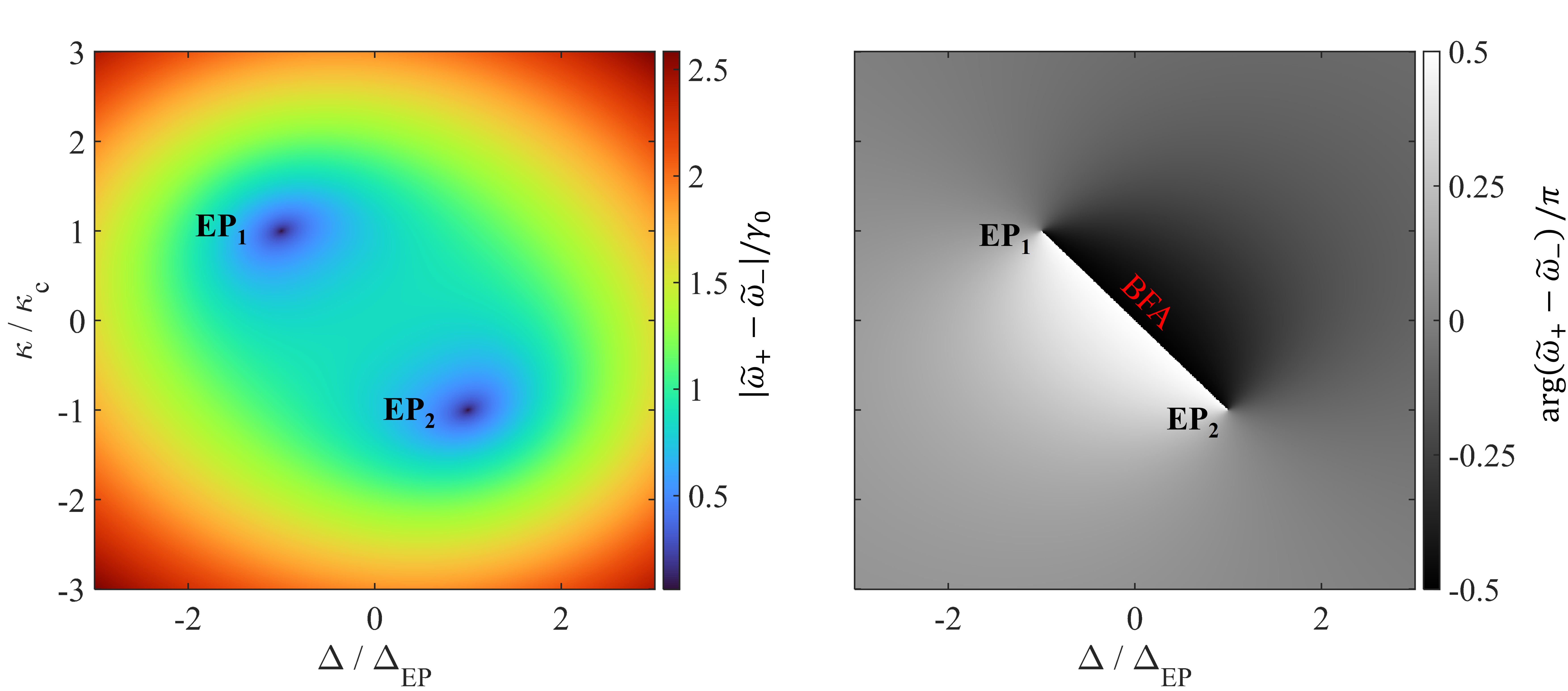}
    \caption{\textbf{Complex splitting in the $(\Delta,\kappa)$ plane.}
    (a) Magnitude $|\tilde{\omega}_+-\tilde{\omega}_-|/\gamma_0$ and (b) phase $\arg(\tilde{\omega}_+-\tilde{\omega}_-)/\pi$ when scanning detuning $\Delta$ and coherent coupling $\kappa$.
    EP$_{1,2}$ are the points where the splitting vanishes.
    The phase is undefined at the EPs and winds by $\pm\pi$ around them (topological charges $\pm\frac12$).
    The phase discontinuity connecting EP$_1$ and EP$_2$ corresponds to a bulk Fermi arc (BFA), defined by $\mathrm{Im}(\tilde{\omega}_+-\tilde{\omega}_-)=0$ (purely real splitting).
    Parameters: $\delta/\gamma_0=0.5$, $\rho=0.8$  and $\phi=0.1\pi$.}
    \label{fig:complexgap}
\end{figure}

In the full two-parameter space $(\Delta,\kappa)$, the EPs appear as isolated zeros of the complex splitting $\tilde{\omega}_+-\tilde{\omega}_-$ [Figure~\ref{fig:complexgap}(a)].
Their topological nature is revealed by the phase of this complex splitting: $\arg(\tilde{\omega}_+-\tilde{\omega}_-)$ is undefined at an EP and winds by $\pm\pi$ upon encircling it once, corresponding to a half-integer topological charge $\pm\frac12$ [Figure~\ref{fig:complexgap}(b)]\cite{Zhen2015,Zhou2018}.
The same phase map also evidences a bulk Fermi arc (BFA) connecting the two EPs, defined by the locus where the splitting is purely real, i.e.\ $\mathrm{Im}(\tilde{\omega}_+-\tilde{\omega}_-)=0$ (equivalently $\arg(\tilde{\omega}_+-\tilde{\omega}_-)=0$ or $\pi$), which acts as a branch cut exchanging the two eigenvalue sheets.

\subsection{Friedrich--Wintgen Interference and Quasi-BICs}
\label{subsec:FW}

The weak-to-strong coupling transition of the previous section is governed by the coherent coupling $\kappa$, the dissipative coupling $\Gamma$ entering only as a renormalization of the threshold [Eq.~(\ref{eq:weak-strong-criterion})]. That same $\Gamma$ now plays the decisive role: it mediates Friedrich--Wintgen (FW) radiative loss exchange between the two modes~\cite{Friedrich1985}. Its effect on the spectrum is most transparent once we note that the eigenvalues of $H_{\mathrm{eff}}$ in Eq.~(\ref{eq:Heff-vsb}) depend on the off-diagonal elements only through their product
\begin{equation}
P \equiv (\kappa-i\Gamma)(\kappa-i\Gamma^{*}).
\label{eq:P-product}
\end{equation}
Consequently, $H_{\mathrm{eff}}$ has exactly the same eigenvalues as the two-mode Hamiltonian $\widetilde{H}_{\mathrm{eff}}$ in which both the coherent coupling $\kappa_{\mathrm{eff}}$ and the dissipative coupling $\sqrt{\gamma_1\gamma_2\,\mu}$ are real and positive,
\begin{equation}
\widetilde{H}_{\mathrm{eff}}=
\begin{pmatrix}
\omega_1-i\gamma_1 & \kappa_{\mathrm{eff}}-i\sqrt{\gamma_1\gamma_2\,\mu}\\[2pt]
\kappa_{\mathrm{eff}}-i\sqrt{\gamma_1\gamma_2\,\mu} & \omega_2-i\gamma_2
\end{pmatrix},
\label{Ht}
\end{equation}
with $\kappa_{\mathrm{eff}}$ and $\mu$ fixed by $P$ through
\begin{equation}
\kappa_{\mathrm{eff}}=\bigl|\mathrm{Re}\sqrt{P}\,\bigr|,
\qquad
\mu=\frac{\bigl(\mathrm{Im}\sqrt{P}\,\bigr)^{2}}{\gamma_1\gamma_2}\in[0,1].
\label{eq:keff-mu}
\end{equation}
 
Equation~\eqref{eq:keff-mu} obtains $\kappa_{\mathrm{eff}}$ and $\mu$ directly from the model through $P$, without separately resolving $\Gamma$ into a modulus $\rho$ and a phase $\phi$, and is therefore valid for arbitrary complex $\Gamma$. The two real parameters partition the off-diagonal coupling into the part that hybridizes the modes and the part that exchanges radiative loss: $\kappa_{\mathrm{eff}}$ is the effective coherent coupling that controls the avoided-crossing gap, while $\mu\in[0,1]$ is the radiation-overlap parameter that controls the strength of the FW loss exchange. Both follow from the bare couplings $(\kappa,\Gamma)$, and $\mu$ in general depends on $\kappa$ as well as on $\Gamma$; the clean separation between hybridization and loss exchange therefore holds in the effective variables $(\kappa_{\mathrm{eff}},\mu)$ rather than in $(\kappa,\Gamma)$. In these variables the weak-to-strong threshold of Eq.~(\ref{eq:weak-strong-criterion}) takes the textbook form $\kappa_{\mathrm{eff}}=|\delta|$, equivalent to $\kappa=\kappa_c$.
 
For real $\Gamma$ ($\mathrm{Im}\,\Gamma\to0$, i.e.\ $\phi=0$ or $\pi$) one has $\kappa_{\mathrm{eff}}\to\kappa$ and $\mu\to\rho^{2}$; more generally $\mu\le\rho^{2}$, with equality only when $\Gamma$ is real. Hence $\rho^{2}$ is the largest overlap available for a given collinearity of the radiation vectors, reached when their top and bottom channels are in phase. In this real-coupling limit $\mu$ admits the intuitive reading of the fraction of radiation that can interfere, from $\mu=0$ for orthogonal radiation patterns to $\mu=1$ for fully collinear, in-phase patterns; for complex $\Gamma$, $\mu$ remains the spectral parameter that governs the loss exchange through Eqs.~\eqref{Ht}--\eqref{eq:keff-mu}.
 
The hybrid linewidths are $\gamma_\pm(\Delta)=-\mathrm{Im}\,\tilde{\omega}_\pm(\Delta)=\gamma_0\mp\mathrm{Im}\,\alpha(\Delta)$, and the FW point, defined by the extremum of the linewidth splitting $\partial_{\Delta}\mathrm{Im}\,\alpha(\Delta)=0$, is located at
\begin{equation}
\Delta_{\mathrm{FW}}=\frac{\kappa_{\mathrm{eff}}\,\delta}{\sqrt{\gamma_1\gamma_2\,\mu}},
\qquad
\gamma_\pm^{\mathrm{FW}}=\gamma_0\pm\sqrt{\delta^{2}+\gamma_1\gamma_2\,\mu}.
\label{FW-linewidth}
\end{equation}
The position $\Delta_{\mathrm{FW}}$ depends on both $\kappa_{\mathrm{eff}}$ and $\mu$, whereas the depth of the suppression, $\gamma_-^{\mathrm{FW}}=\gamma_0-\sqrt{\delta^{2}+\gamma_1\gamma_2\,\mu}$, is independent of $\kappa_{\mathrm{eff}}$ and is set by the overlap $\mu$ and the bare loss imbalance $\delta$ alone.
 
A true FW bound state in the continuum appears when the lower hybrid linewidth vanishes,
\begin{equation}
\gamma_-^{\mathrm{FW}}=0
\quad\Longleftrightarrow\quad
\mu=1.
\label{FW-BIC}
\end{equation}
Since $\mu\le\rho^{2}\le1$, this requires $\rho=1$ together with in-phase top/bottom channels, i.e.\ radiation vectors that are collinear in the full radiation-channel space with the appropriate global phase, so that the radiative amplitudes of one hybrid state cancel simultaneously in all open channels~\cite{Fan2003,WonjooSuh2004}. When several diffraction orders are open, this is stronger than matching a single outgoing polarization. More generally, as $\mu$ approaches unity, destructive interference strongly suppresses one hybrid linewidth while enhancing the other, producing a quasi-BIC whose robustness is controlled by $\mu$ rather than by the coherent gap~\cite{Koshelev2018,Blanchard2016}. Quasi-BIC behaviour can therefore occur in the weak-coupling regime, at the EP, or in the strong-coupling regime, provided vertical-symmetry breaking aligns the radiation channels efficiently~\cite{Gowadzka2021,Cerjan2019}.
 
This is illustrated in Figs.~\ref{fig:transition_noQuasiBIC}(a)--(f), where in both rows $\delta/\gamma_0=0.5$ and the bare coherent coupling is swept through the weak--EP--strong sequence, $\kappa=0.6\kappa_c,\ \kappa_c,\ 3\kappa_c$. For Figs.~\ref{fig:transition_noQuasiBIC}(a)--(c) the radiation overlap is moderate ($\rho=0.8$, $\phi=0.3\pi$), giving $\mu\simeq0.5$ at weak coupling and decreasing toward strong coupling, so the linewidths merely exchange without a deep minimum. For Figs.~\ref{fig:transition_noQuasiBIC}(d)--(f) the radiation vectors are nearly collinear ($\rho=0.98$, $\phi=0.05\pi$), giving $\mu\simeq0.95$, and the same loss-exchange mechanism now produces a pronounced quasi-BIC minimum. Thus the avoided-crossing gap is set by $\kappa_{\mathrm{eff}}$, whereas the depth of the linewidth suppression---and hence the quasi-BIC character---is set by the overlap $\mu$ (at fixed bare imbalance $\delta$).
 \begin{figure}[ht!]
    \centering
    \includegraphics[width=0.62\linewidth]{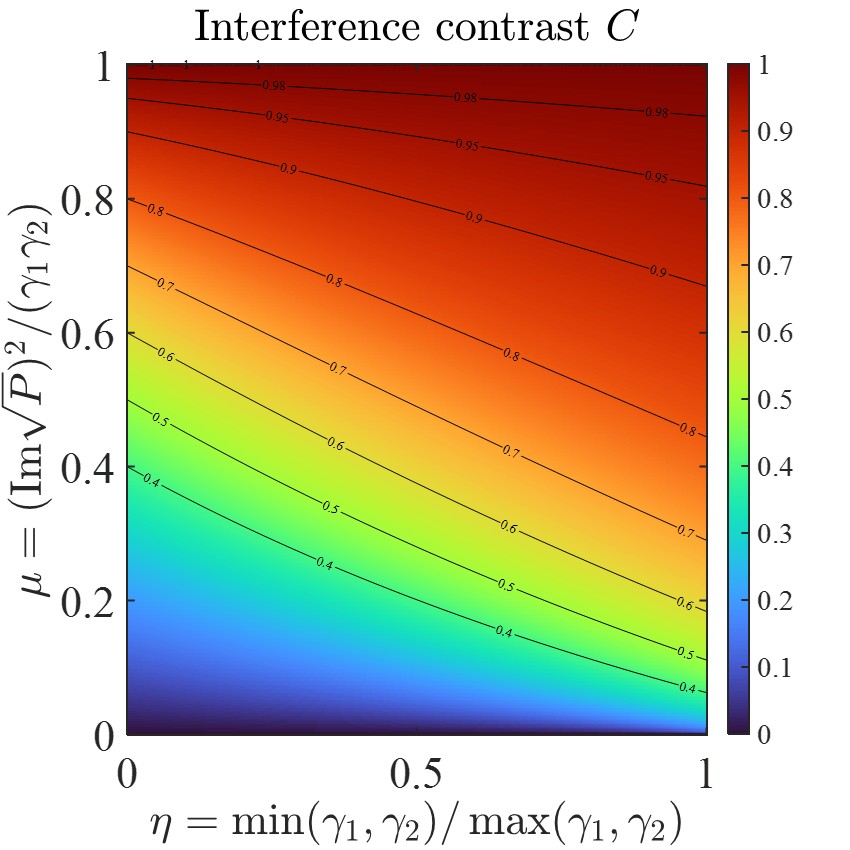}
    \caption{\textbf{Map of FW interference contrast.}
    Color map of $C(\eta,\mu)$ defined in Eq.~\eqref{eq:C-eta-phi}, showing how loss-exchange efficiency depends on the loss ratio $\eta=\min(\gamma_1,\gamma_2)/\max(\gamma_1,\gamma_2)$ and the radiation-overlap parameter $\mu=(\mathrm{Im}\sqrt{P})^{2}/(\gamma_1\gamma_2)$ defined in Eq.~\eqref{eq:keff-mu}.
    High contrast is most easily obtained when the two uncoupled modes have comparable radiative losses ($\eta\approx 1$), in which case the tolerance to imperfect overlap is enhanced.}
    \label{fig:contrast}
\end{figure}
To quantify how efficiently radiative loss is exchanged at $\Delta=\Delta_{\mathrm{FW}}$, we introduce the normalized FW contrast
\begin{equation}
C = 1 - \frac{\gamma_-^{\mathrm{FW}}}{\gamma_+^{\mathrm{FW}}}\,\frac{\max(\gamma_1,\gamma_2)}{\min(\gamma_1,\gamma_2)},
\label{C-def}
\end{equation}
with $\gamma_\pm^{\mathrm{FW}}=\gamma_\pm(\Delta_{\mathrm{FW}})$. For $\mu=0$ there is no dissipative loss exchange: the hybrid linewidths at the FW point reduce to the bare ones, $\gamma_-^{\mathrm{FW}}=\min(\gamma_1,\gamma_2)$ and $\gamma_+^{\mathrm{FW}}=\max(\gamma_1,\gamma_2)$, so $C=0$. By contrast, $C\to1$ only when FW interference transfers essentially all radiative loss to one hybrid branch, leaving a deeply suppressed, quasi-BIC branch. This normalization matters because a bare asymmetry measure such as $S=(\gamma_+^{\mathrm{FW}}-\gamma_-^{\mathrm{FW}})/(\gamma_+^{\mathrm{FW}}+\gamma_-^{\mathrm{FW}})$ can be large even at $\mu=0$, simply because $\gamma_1\neq\gamma_2$, and thus conflates genuine FW exchange with the intrinsic loss imbalance.
 
Inserting Eq.~\eqref{FW-linewidth} into Eq.~\eqref{C-def}, with the loss ratio $\eta=\min(\gamma_1,\gamma_2)/\max(\gamma_1,\gamma_2)$ and the overlap $\mu$ of Eq.~\eqref{eq:keff-mu}, yields the closed form
\begin{equation}
C(\eta,\mu)= 1 - \frac{1}{\eta}\,\frac{(1+\eta)-\sqrt{(1-\eta)^2+4\eta\mu}}{(1+\eta)+\sqrt{(1-\eta)^2+4\eta\mu}},
\label{eq:C-eta-phi}
\end{equation}
which depends only on $\eta$ and $\mu$ and is independent of the absolute loss scale; because $\mu$ is defined through $P$, it holds for arbitrary complex $\Gamma$. Accordingly $C$ vanishes at $\mu=0$ and approaches unity as $\mu\to1$, where the lower linewidth is strongly suppressed, $\gamma_-^{\mathrm{FW}}\ll\gamma_0$, giving the experimentally relevant quasi-BIC regime; the ideal FW-BIC is recovered at $\mu=1$. Figure~\ref{fig:contrast} maps $C(\eta,\mu)$ and shows that near-unity loss exchange ($C\to1$) is reached for $\mu\to1$, and that this high-contrast regime is markedly more tolerant of imperfect overlap when the bare losses are comparable ($\eta\simeq1$): balancing the linewidths enlarges the parameter window for deep suppression without requiring exact cancellation.

\subsection{Unidirectional Guided Resonances}
\label{sec:UGR}

The two non-Hermitian singularities discussed so far---EPs and
Friedrich--Wintgen (quasi-)BICs---are both encoded in the total hybrid
linewidth $\gamma_{\mathbf c}$. The radiation vector
$\mathbf K_{\mathbf c}=c_1\mathbf K_1+c_2\mathbf K_2$ introduced in Sec.~\ref{sec:General_theory},
however, carries direction-resolved information that the total decay rate discards.
A natural extension within the same Hamiltonian is therefore to ask under what
conditions an eigenmode of $H_{\mathrm{eff}}$ becomes dark to one half-space while
remaining radiative toward the other---i.e.\ a UGR ~\cite{Zhou2016,Yin2020,Yin2023,Wang2024,Lee2024,Lee2025,Zhuang2024,Yuan2026,Choi2025,Maksimov2025,YuanLu2025}.
Such single-sided radiation is the central design target for high-efficiency grating
couplers, and directional
antennas, and has been independently established as a phenomenon of vertical-symmetry
breaking in both single-layer
~\cite{Yin2023,Lee2024,Lee2025,Maksimov2025} and multilayer
~\cite{Zhuang2024,Yuan2026,Choi2025} photonic crystal slabs. Our purpose here is not to
introduce UGRs, but to show that they can emerge from exactly the same two-mode
non-Hermitian Hamiltonian Eq.~\eqref{eq:Heff-vsb} that governs the weak-to-strong
coupling transition and the FW quasi-BIC, with a clean topological characterization
that places UGRs on the same footing as the EP and the FW BIC.

Throughout this subsection we work in the sub-diffractive regime with a single open radiation channel per side---the zeroth diffraction order in the dominant polarization, as in the case studies below---so that each $K_m^{s}$ is a scalar amplitude, the direction-resolved overlaps have unit modulus, $\rho_t=\rho_b=1$, and the amplitude ratio $\beta_s\equiv|K_2^{\,s}|/|K_1^{\,s}|=\sqrt{\gamma_2^{\,s}/\gamma_1^{\,s}}$ and phase $\phi_s=\arg(K_2^{\,s})-\arg(K_1^{\,s})$ are well defined. The spectral results---the EP and the FW quasi-BIC---hold for the full multi-channel radiation vector; only the explicit UGR condition below relies on this single-channel reduction, which is also what renders one-sided radiation cancellation a codimension-one rather than codimension-two condition~\cite{YuanLu2025}.

We first discuss the direction-resolved hybrid radiation. For any hybrid eigenmode of $H_{\mathrm{eff}}$ with coefficient vector
$\mathbf c=(c_1,c_2)^T$, the side-resolved radiation amplitudes read
\begin{equation}
A_{\mathbf c}^{s}=c_1\,K_1^{s}+c_2\,K_2^{s},
\qquad s=t,b,
\label{eq:UGR-Apm}
\end{equation}
with corresponding direction-resolved decay rates
$\gamma_{\mathbf c}^{s}=\tfrac12|A_{\mathbf c}^{s}|^2$ and
$\gamma_{\mathbf c}=\gamma_{\mathbf c}^{t}+\gamma_{\mathbf c}^{b}$. A convenient
scalar summary is the directional contrast
\begin{equation}
\mathcal{D}_{\mathbf c}=
\frac{\gamma_{\mathbf c}^{\,t}-\gamma_{\mathbf c}^{\,b}}
     {\gamma_{\mathbf c}^{\,t}+\gamma_{\mathbf c}^{\,b}}\in[-1,1],
\label{eq:UGR-contrast}
\end{equation}
which vanishes in the vertically symmetric reference and reaches $\pm 1$ when emission
is purely unidirectional---the operational definition of UGR
~\cite{Zhou2016,Yin2023,Lee2024}. The total decay rate $\gamma_{\mathbf c}$ and the contrast $\mathcal D_{\mathbf c}$ are simply the sum and the normalized difference of the same side-resolved rates $\gamma_{\mathbf c}^{t,b}$: the sum is the linewidth whose coalescence and minimum define the EP and the FW (quasi-)BIC of Sec.~\ref{subsec:FW}, while the difference defines the (quasi-)UGR. EPs, (quasi-)BICs and (quasi-)UGRs are thus three readings of one object.

The UGR condition can be derived from the eigenvector ratio of $H_{\mathrm{eff}}$. Eigenvectors of $H_{\mathrm{eff}}$ from Eq.~\eqref{eq:Heff-vsb} obey
\begin{equation}
\frac{c_1^\pm}{c_2^\pm}=
-\frac{\kappa-i\Gamma}{(\Delta-i\delta)\mp\alpha(\Delta)} .
\label{eq:UGR-eigratio}
\end{equation}
Setting $A_{\mathbf c}^{\,s}=0$ for $s\in\{t,b\}$ yields the UGR condition
\begin{equation}
\left.\frac{c_1^\pm}{c_2^\pm}\right|_{\mathrm{UGR}_s}
= -\,\frac{K_2^{\,s}}{K_1^{\,s}}
= -\,\beta_s\,e^{i\phi_s},
\quad s=t,b,
\label{eq:UGR-condition}
\end{equation}
where $\beta_s=|K_2^{\,s}|/|K_1^{\,s}|=\sqrt{\gamma_2^{\,s}/\gamma_1^{\,s}}$ is the
modulus ratio of the parent-mode radiation amplitudes on the side $s$, and
$\phi_s=\arg(K_2^{\,s})-\arg(K_1^{\,s})$ is the corresponding relative radiative phase.
In the vertically symmetric reference structure, parity enforces $\beta_t=\beta_b$ and
$\phi_t-\phi_b=\pi$, and the off-diagonal couplings vanish ($\kappa=\Gamma=0$). The
eigenmodes are then pinned to the pure even/odd parity states, each of which radiates
with equal weight into the top and bottom half-spaces ($\mathcal{D}_{\mathbf c}=0$);
moreover, the top-dark and bottom-dark conditions of Eq.~\eqref{eq:UGR-condition} differ
only by a sign and cannot be met selectively. Consequently, no UGR exists in the
vertically symmetric limit. Vertical-symmetry breaking is therefore essential: it
simultaneously activates the coherent and dissipative couplings---enabling genuine
hybrid superpositions of the two parent modes---and lifts the parity constraints by
independently retuning $\beta_t/\beta_b$ and $\phi_t-\phi_b$, so that one hybrid mode
can become dark on a single side while remaining radiative on the other, generically
opening a UGR window. The same mechanism was identified for vertically-asymmetric single-layer
slabs in Ref.~\cite{Yin2023}, for L-shaped and zero-contrast gratings in
Refs.~\cite{Lee2024,Lee2025}, and for hetero-bilayer slabs in Ref.~\cite{Yuan2026}.

Making this explicit, the side-resolved decay rate of Eq.~\eqref{eq:UGR-Apm} reads
\begin{equation}
\gamma_{\mathbf c}^{s}=|c_1|^2\gamma_1^{s}+|c_2|^2\gamma_2^{s}+2\sqrt{\gamma_1^{s}\gamma_2^{s}}\,\mathrm{Re}\!\left(c_1c_2^{*}\,e^{-i\phi_s}\right),\quad s=t,b,
\label{eq:gamma-side}
\end{equation}
{so that inserting the eigenvector ratio Eq.~\eqref{eq:UGR-eigratio} makes both $\gamma_{\mathbf c}$ and $\mathcal D_{\mathbf c}(\Delta)$ explicit in the model parameters. It is worth stating what direction resolution costs in model input. The eigenvalues---hence the EP and the FW quasi-BIC---depend on the off-diagonal coupling only through $P$, i.e.\ through $\kappa_{\mathrm{eff}}$ and $\mu$ [Eq.~\eqref{eq:keff-mu}]; equivalently, the spectrum sees only the coherent \emph{sum} of the top and bottom radiative overlaps that constitutes $\Gamma$ [Eq.~\eqref{eq:rho-top-bottom}]. The contrast $\mathcal D_{\mathbf c}$ instead depends on the two side contributions \emph{separately}, and is therefore sensitive to the per-side quantities that this sum leaves free---the amplitude ratio $\beta_t/\beta_b$ and the phase difference $\phi_t-\phi_b$, released from their parity values ($1$ and $\pi$) by vertical-symmetry breaking. The EP and the FW quasi-BIC are thus \emph{spectral}, fixed by $\kappa_{\mathrm{eff}}$ and $\mu$, whereas the UGR is a strictly finer, direction-resolved property requiring the full side-resolved radiation vectors $\{K_m^{t,b}\}$ beyond $P$.

Combining Eqs.~\eqref{eq:UGR-eigratio}--\eqref{eq:UGR-condition} yields
\begin{equation}
\kappa-i\Gamma
=\beta_s\,e^{i\phi_s}\bigl[(\Delta-i\delta)\mp\alpha(\Delta)\bigr],
\label{eq:UGR-implicit}
\end{equation}
which fixes, for a given side $s$, the relation that the model and radiation
parameters must satisfy for one hybrid mode to go dark there. The two control knobs
in our problem are the detuning $\Delta$, set by the in-plane wavevector
$k_\parallel$, and a structural symmetry-breaking parameter that tunes $\kappa$ and the radiation overlap.
Equation ~\eqref{eq:UGR-implicit} is formally a complex equation, which by naive counting would make perfect one-sided cancellation a codimension-two condition - an isolated point in the two-parameter plane. For a single open radiation channel per side, however, the analytic structure of the resonant scattering problem removes one of these two conditions, so that UGRs are in fact codimension-one objects, as established rigorously in Ref~\cite{YuanLu2025}: they trace continuous curves
---ridges--- in the two-parameter plane, and a continuous family of UGRs spawns from
each BIC upon tuning a single parameter. Such robustness will be illustrated in the two case studies considered in this work, as discussed below.


\section{Case Study I: Vertical symmetry breaking with superstrate versus substrate and partial etching}
\label{sec:numerics}
\subsection{Numerical Demonstration}
We now illustrate the general framework using a realistic and fabrication-friendly square-lattice photonic crystal slab, which also serves as the platform for the experiments discussed below. The structure consists of a dielectric slab with refractive index $n_{\mathrm{slab}}=2.03$ and thickness $h=130~\mathrm{nm}$, patterned with a square lattice of circular holes with period $a=354~\mathrm{nm}$ and diameter $d=160~\mathrm{nm}$, on a substrate with refractive index $n_{\mathrm{sub}}=1.46$. For the Rigorous Coupled-Wave Analysis (RCWA)~\cite{Liu2012} absorption spectra, a weak numerical probe loss was introduced by assigning an imaginary refractive-index component of $10^{-7}$ to the slab material. This artificial absorption channel is used only to reveal the resonant features in the absorption map and is sufficiently small that the modal dispersion remains essentially unchanged. The slab index is representative of common dielectric materials such as SiN$_x$ or TiO$_2$, while the substrate index is typical of SiO$_2$. Throughout this section, we consider $s$-polarized excitation and detection, with the electric field oriented along $y$, and scan the in-plane wavevector along the $\Gamma X$ direction. In the vertically symmetric limit, the two relevant leaky resonances originate from TE-like and TM-like slab modes with opposite vertical parity. They can therefore be regarded as odd-like and even-like resonances whose mutual coupling is forbidden by symmetry. Angle-resolved absorption spectra are calculated using RCWA simulations, while the corresponding complex eigenfrequencies are obtained from Finite Element Method (FEM) eigenmode simulations using COMSOL Multiphysics.

\begin{figure*}[ht!]
    \centering
    \includegraphics[width=1\linewidth]{asymmetrization_v3.jpg}
    \caption{\textbf{Vertical-symmetry-breaking pathways from mode crossing to avoided crossing and quasi-BIC formation.}
    First row: schematic illustrations of the vertically symmetric reference structure and three vertical-symmetry-breaking configurations. 
    Second row: angle-resolved $s$-polarized absorption spectra along the $\Gamma X$ direction, plotted as $\log_{10}(A)$. A weak numerical probe loss, corresponding to an imaginary refractive-index component of $10^{-7}$ in the slab material, was included in the RCWA absorption calculations.
    Third and fourth rows: corresponding eigenfrequency dispersions and radiative linewidths $\gamma$ of the two hybrid branches obtained from FEM eigenmode simulations. 
    (a) In the vertically symmetric reference structure, identical superstrate and substrate environments preserve mirror symmetry. The two resonances retain opposite vertical parity and therefore cross without appreciable hybridization. 
    (b) In the index-matching-liquid configuration, the superstrate refractive index is set to $n_{\mathrm{sup}}=1.426$, while the slab geometry is unchanged. This weak environmental asymmetry preserves a crossing-like dispersion, and the linewidths remain separated, indicating that the system remains in the weak-coupling regime. 
    (c) In the fully etched structure, the vertical asymmetry is increased by patterning through the full slab thickness. The two resonances undergo a clear avoided crossing, indicating strong coherent hybridization, together with redistribution of the radiative linewidths. 
    (d) In the partially etched structure with etching ratio of 0.5, a residual unpatterned layer remains below the patterned region. This geometry provides additional control over the relative radiation phase and overlap, producing an avoided crossing together with strong linewidth suppression on one hybrid branch. The linewidth minimum corresponds to a Friedrich--Wintgen quasi-BIC. 
    Geometrical and material parameters are given in the main text.}
    \label{fig:simul_trans}
\end{figure*}

Figure~\ref{fig:simul_trans} compares a vertically symmetric reference structure with three representative pathways for breaking vertical symmetry, showing how the system evolves from symmetry-protected crossing modes to hybridized leaky resonances. In the vertically symmetric reference structure [Figure~\ref{fig:simul_trans}(a)], the superstrate and substrate are identical, so the two resonances remain orthogonal under the vertical mirror operation. Consequently, the angle-resolved absorption map shows a real-frequency crossing, and the extracted eigenfrequencies cross without any linewidth exchange. Introducing an index-matching liquid with refractive index $n_{\mathrm{sup}}=1.426$ [Figure~\ref{fig:simul_trans}(b)] weakly breaks vertical symmetry through the dielectric environment rather than through the photonic crystal geometry itself. In this case, the absorption spectrum and the extracted eigenfrequencies still show a crossing-like dispersion. The linewidths vary strongly but remain separated across the crossing region, without a complete linewidth exchange. This behavior indicates that the induced coherent coupling remains below the weak-to-strong coupling threshold. The system therefore remains in the weak-coupling regime, although the vertical asymmetry can still perturb the radiative decay rates locally near the crossing. Later in this section, we show that by carefully tuning the superstrate refractive index around this value, the system can be continuously driven to the exceptional-point condition at the boundary between weak and strong coupling. A stronger structural asymmetry is obtained in the fully etched slab [Figure~\ref{fig:simul_trans}(c)], where the air holes extend through the full slab thickness. In this case, the real-frequency crossing is lifted and replaced by an avoided crossing, demonstrating the onset of strong coherent hybridization between the two resonances. The accompanying linewidth redistribution confirms that the same vertical-symmetry-breaking perturbation also mixes the outgoing radiation channels. A qualitatively new regime emerges in Figure~\ref{fig:simul_trans}(d), where vertical-symmetry breaking is further strengthened by partial etching.
While the frequency splitting remains of the same order (consistent with a comparable $\kappa$), one hybrid branch now exhibits a pronounced  linewidth dip in the vicinity of the avoided crossing, i.e.\ a quasi-BIC. As discussed in the following, this quasi-BIC feature persists over a broad range of partial etching depths.
This behavior indicates highly efficient destructive interference of the radiative channels: perfect cancellation is not required for a strong suppression, and a narrow resonance can persist over a finite parameter window.

\begin{figure*}[ht!]
    \centering
    \includegraphics[width=1\linewidth]{ep.jpg}
    \caption{\textbf{Numerical demonstration of the weak-to-strong coupling transition, with an EP emerging at the boundary.} The first and second rows show the real and imaginary parts of the complex eigenfrequencies of the PhC slab, calculated by FEM eigenmode simulations for different superstrate refractive indices. Panels (a)--(c) illustrate the continuous transition from the weak-coupling regime to the strong-coupling regime through an EP: in (a), the real parts cross while the imaginary parts remain separated; in (b), both the real and imaginary parts coalesce at the EP; and in (c), the real parts undergo an avoided crossing while the imaginary parts exchange between the two branches. Panels (d) and (e) show, respectively, the magnitude and phase of the complex eigenfrequency gap when both the in-plane wavevector and the superstrate refractive index are varied. The EP is identified by the vanishing complex gap in panel (d) and by a $\pm\pi$ phase winding in panel (e), corresponding to a half-integer topological charge of $\pm 1/2$. The three white dashed lines labeled A, B, and C in panels (d) and (e) indicate the three selected superstrate refractive indices used in panels (a), (b), and (c), respectively. Geometrical and material parameters are given in the main text.}
    \label{fig:simul_fem}
\end{figure*}
Figure~\ref{fig:simul_fem} presents zoom-in full-wave eigenfrequency simulations of the photonic crystal slab obtained by the FEM. By treating the superstrate refractive index $n$ as a symmetry-breaking parameter, 
the results capture the continuous evolution from weak coupling to the strong-coupling 
regime. In Figure~\ref{fig:simul_fem}(a), a real-frequency crossing persists at $n = 1.4259$ while the linewidths remain split, characterizing the weak-coupling regime. At the boundary of weak-to-strong coupling ($n = 1.425537$), the system reaches the EP threshold shown in Figure~\ref{fig:simul_fem}(b), where the hybrid-mode frequencies and decay rates coalesce. Further increasing the vertical asymmetry to $n = 1.4251$ drives the system into the strong-coupling regime depicted in Figure~\ref{fig:simul_fem}(c). In this regime, the real frequencies undergo a clear avoided crossing, maintaining a finite frequency gap for all values of $k_{\parallel}$, while the radiative linewidths cross to facilitate the exchange of loss between the hybrid modes. Figures~\ref{fig:simul_fem}(d) and \ref{fig:simul_fem}(e) confirm the topological nature of this singularity; the complex frequency gap vanishes at a localized point in the $(k_{\parallel}, n)$ parameter space, around which the phase winds by $\pm\pi$, indicating a half-integer topological charge of $\pm 1/2$.

\begin{figure*}[ht!]
    \centering
    \includegraphics[width=1\linewidth]{Simulation_contrast_bic_urg.jpg}
    \caption{\textbf{Engineering Friedrich--Wintgen quasi-BICs and direction-resolved radiation through partial etching.}
    Effective two-mode parameters and radiation properties of the partially etched PhC slab as a function of etching ratio, defined as the patterned depth divided by the total slab thickness.
    (a)--(c) Mode energies, radiative linewidths, and quality factors for a representative low-etching-ratio case of $0.2$, where near-unity radiation overlap $\mu$ produces a pronounced linewidth minimum and strong quality-factor enhancement.
    (d)--(f) Same quantities for a larger etching ratio of $0.8$, where reduced radiation overlap weakens the linewidth suppression and lowers the quality-factor enhancement. Symbols denote extracted numerical data for the upper and lower branches, while solid curves show the analytical two-mode fit.
    (g) Effective coherent coupling strength $\kappa_{\mathrm{eff}}=|\mathrm{Re}\sqrt{P}|$ as a function of etching ratio, where $P=(\kappa-i\Gamma)(\kappa-i\Gamma^{*})$.
    (h) Radiation-overlap parameter $\mu=(\mathrm{Im}\sqrt{P})^{2}/(\gamma_1\gamma_2)$, obtained from the same two-mode fit.
    (i) Normalized Friedrich--Wintgen interference contrast $C$ evaluated at the FW point. The shaded region marks the quasi-BIC plateau, where near-unity $C$ indicates efficient radiative loss exchange.
    (j) FEM-calculated quality-factor map, $\log_{10}(Q)$, in the $(k_xa/\pi,\mathrm{etching~ratio})$ parameter space, showing a high-$Q$ ridge associated with quasi-BIC formation.
    (k) FEM-calculated directional contrast $\mathcal{D}_{\mathbf c}$ in the same parameter space. The dashed curve indicates the trajectory of maximum quality factor. The largest $|\mathcal{D}_{\mathbf c}|$ does not exactly coincide with the quasi-BIC ridge, demonstrating that quasi-BIC formation and UGR-like one-sided radiation cancellation correspond to different constraints on the same hybrid radiation vector.}
    \label{fig:simul_contrast}
\end{figure*}
To make the role of partial etching quantitative, Figure~\ref{fig:simul_contrast} analyzes how the etching ratio controls both Friedrich--Wintgen quasi-BIC formation and direction-resolved radiation in the single-layer PhC slab, where the etching ratio is defined as the patterned depth divided by the total slab thickness. All other geometrical and material parameters are kept identical to the partially etched design described above.

Representative two-mode fits are shown first in Figures~\ref{fig:simul_contrast}(a)--\ref{fig:simul_contrast}(f). At a low etching ratio of $0.2$ [Figures~\ref{fig:simul_contrast}(a)--\ref{fig:simul_contrast}(c)], the analytical model reproduces the extracted mode energies, radiative linewidths, and quality factors. The lower branch exhibits a deep linewidth minimum near the Friedrich--Wintgen point, accompanied by a strong quality-factor enhancement, confirming quasi-BIC formation through efficient radiative loss exchange. By contrast, at a larger etching ratio of $0.8$ [Figures~\ref{fig:simul_contrast}(d)--\ref{fig:simul_contrast}(f)], the linewidth suppression becomes much weaker and the quality-factor enhancement is strongly reduced. These results show that coherent hybridization alone is not sufficient to produce a deep quasi-BIC; efficient Friedrich--Wintgen suppression also requires favorable far-field radiation overlap.

The effective two-mode parameters extracted from the partially etched design are summarized in Figures~\ref{fig:simul_contrast}(g)--\ref{fig:simul_contrast}(i). The effective coherent coupling strength $\kappa_{\mathrm{eff}}$ [Figure~\ref{fig:simul_contrast}(g)] varies strongly with etching ratio, confirming that partial etching directly tunes the near-field hybridization and therefore the avoided-crossing gap. The radiation-overlap parameter $\mu$ [Figure~\ref{fig:simul_contrast}(h)] remains close to unity for small and moderate etching ratios, indicating nearly phase-matched radiation into the same outgoing channel, but decreases rapidly at larger etching ratios as the relative radiation phase becomes mismatched. This trend directly controls the normalized Friedrich--Wintgen contrast $C$ [Figure~\ref{fig:simul_contrast}(i)], which forms a near-unity plateau for etching ratios below approximately $0.5$. We refer to this interval as the quasi-BIC plateau, where efficient radiative loss exchange can strongly suppress the linewidth of one hybrid branch.

Figures~\ref{fig:simul_contrast}(j) and \ref{fig:simul_contrast}(k) extend this analysis from total radiative loss to direction-resolved radiation using FEM eigenmode simulations. Figure~\ref{fig:simul_contrast}(j) shows the quality-factor map, $\log_{10}(Q)$, in the $(k_xa/\pi,\mathrm{etching~ratio})$ parameter space. A high-$Q$ ridge appears near the hybridization region, confirming robust quasi-BIC formation through Friedrich--Wintgen interference. Figure~\ref{fig:simul_contrast}(k) shows the corresponding directional contrast $\mathcal{D}_{\mathbf c}$ in the same parameter space, extracted from the upward and downward radiative power of the FEM eigenmodes. The dashed curve indicates the trajectory of maximum quality factor. Importantly, the largest $|\mathcal{D}_{\mathbf c}|$ occurs near, but does not exactly coincide with, the quasi-BIC trajectory. This separation demonstrates that quasi-BIC formation and UGR-like one-sided radiation cancellation are related but distinct interference conditions: the quasi-BIC suppresses the total radiative decay, whereas the UGR-like condition suppresses radiation preferentially into one half-space while maintaining finite radiation into the opposite half-space. Therefore, partial etching provides simultaneous control over near-field coherent coupling, far-field radiative loss exchange, and direction-resolved emission.

\subsection{Experimental realization}
\label{sec:exp}
To validate experimentally the vertical-symmetry-breaking mechanism for Friedrich--Wintgen interference and quasi-BIC formation, we fabricate a square-lattice SiN$_x$-on-SiO$_2$ PhC slab matching the design discussed in Sec.~\ref{sec:numerics}. A $130\,$nm layer of SiN$_x$ is first deposited on a fused-silica substrate by Plasma-Enhanced Chemical Vapor Deposition (PECVD). PMMA resist is then applied by spin-coating, and electron-beam lithography is used to define an $80\times80\,\mu\text{m}^2$ square-lattice pattern with period $a=354\,$nm and hole diameter $d=160\,$nm. The pattern is transferred into the SiN$_x$ layer by Reactive Ion Etching (RIE) using a CH$_4$/O$_2$ gas mixture, with the etch carefully controlled to achieve a partial etch depth corresponding to $\epsilon=43\%$ of the total slab thickness, thereby introducing the designed degree of vertical asymmetry. The remaining PMMA resist is subsequently stripped by O$_2$ plasma exposure. Angle-resolved reflectivity is then used to measure the energy--momentum dispersion of the Bloch resonances~\cite{Cueff2024}.

\begin{figure*}[ht!]
    \centering
    \includegraphics[width=1\linewidth]{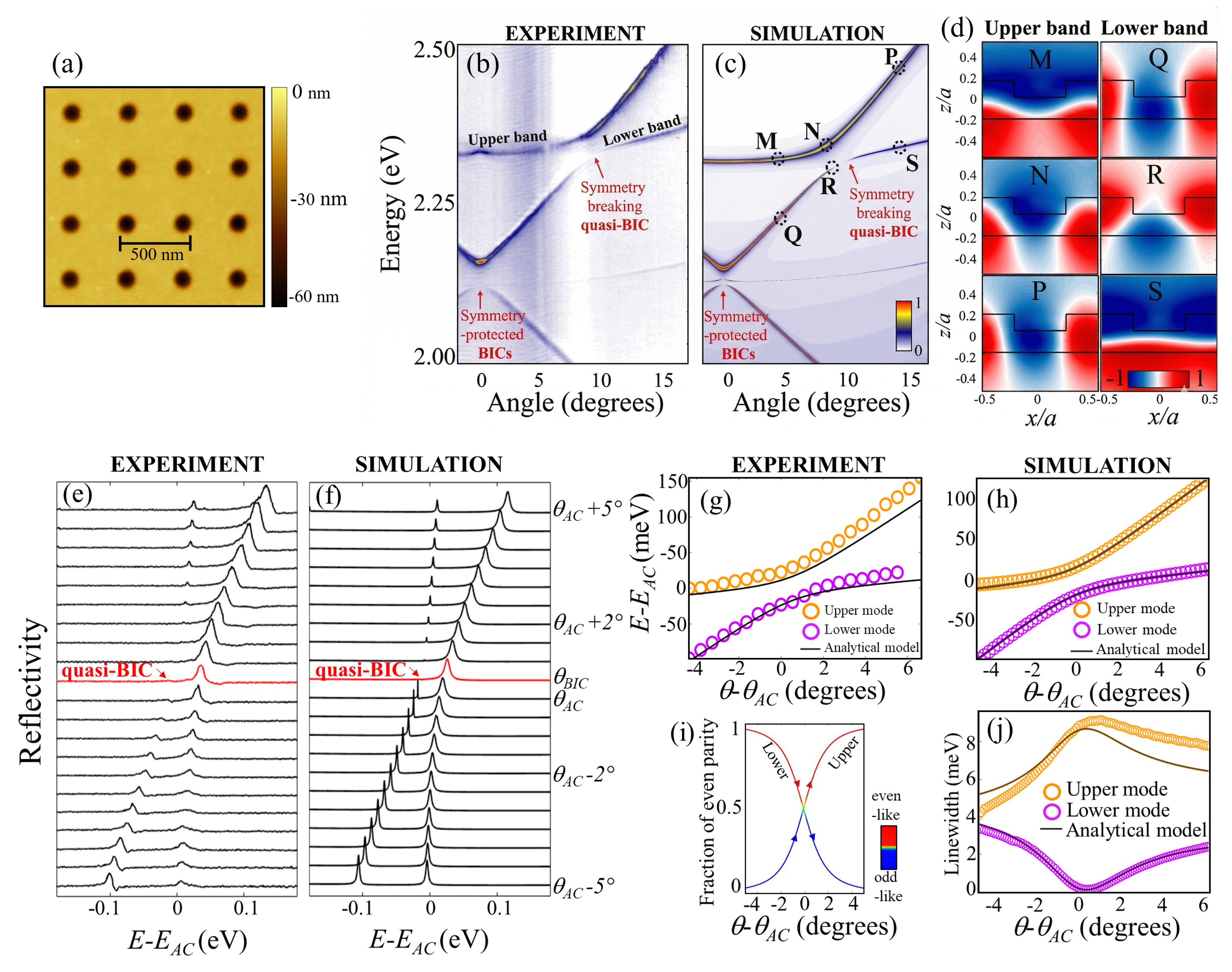}
    \caption{\textbf{Experimental observation and quantitative validation of the symmetry-breaking quasi-BIC.}
    (a) AFM image of the fabricated partially etched square-lattice PhC slab. Scale bar: $500~\mathrm{nm}$. 
    (b) Measured $s$-polarized angle-resolved reflectivity spectrum along $\Gamma X$, showing $\Gamma$-point symmetry-protected BICs and an off-$\Gamma$ symmetry-breaking quasi-BIC near the avoided-crossing region. 
    (c) RCWA simulation of the same structure, reproducing the measured dispersion and the quasi-BIC feature. 
    (d) Representative real-space maps of $E_y$ at the labeled points M--S in panel (c), illustrating the exchange between even-like and odd-like modal profiles across the hybridization region. 
    (e) Experimental reflectivity spectra for incident angles around the avoided crossing. The quasi-BIC spectrum is highlighted in red. 
    (f) Corresponding simulated reflectivity spectra. 
    (g) Mode energies extracted from the experimental spectra in panel (e), together with the analytical two-mode model. 
    (h) Mode energies extracted from the simulated spectra in panel (f), together with the same analytical model. 
    (i) Calculated fraction of even parity for the two hybrid branches, showing parity exchange across the avoided crossing. 
    (j) Simulated radiative linewidths extracted near the hybridization region, together with the analytical two-mode model. The linewidth minimum on one hybrid branch confirms Friedrich--Wintgen loss exchange and quasi-BIC formation. The solid lines in panels (g), (h), and (j) are analytical fits to the real parts in panels (g) and (h) and the imaginary part in panel (j), obtained using a single set of parameters: $\kappa_{\mathrm{eff}}= 17.7$\,meV, $\gamma_e = 5.3$\,meV, $\gamma_o = 3.5$\,meV, $\alpha_e = 1.2$\,eV/rad, and $\alpha_o = 0.12$\,eV/rad.}
    \label{fig:exp}
\end{figure*}

Figure~\ref{fig:exp} summarizes both the experimental observation and the quantitative validation of the symmetry-breaking quasi-BIC. The AFM image in Figure~\ref{fig:exp}(a) confirms the fabricated square-lattice hole array. The measured $s$-polarized angle-resolved reflectivity spectrum in 
Figure~\ref{fig:exp}(b) reveals several guided resonances, including 
symmetry-protected BICs at the $\Gamma$ point and an additional 
linewidth-suppressed feature away from $\Gamma$. This off-$\Gamma$ feature 
appears near the avoided crossing between a relatively flat upper band and a 
more dispersive lower band, and is identified as a symmetry-breaking quasi-BIC. 
The corresponding RCWA simulation in Figure~\ref{fig:exp}(c) reproduces the 
measured dispersion and the quasi-BIC feature, confirming that the observed 
linewidth suppression originates from the designed vertically asymmetric PhC slab. 
An analogous off-$\Gamma$ quasi-BIC is also observed in the $p$-polarized channel 
of the same structure, as demonstrated in the Supplementary Material.

The modal profiles in Figure~\ref{fig:exp}(d) clarify the physical origin of the quasi-BIC. Away from the hybridization region, the two branches recover predominantly even-like or odd-like field profiles. Near the avoided crossing, however, the two modes strongly hybridize and exchange their parity character. This parity mixing enables the two resonances to radiate into overlapping far-field channels, providing the condition for Friedrich--Wintgen destructive interference. The off-$\Gamma$ quasi-BIC therefore differs from the $\Gamma$-point symmetry-protected BICs: it is not protected by an in-plane symmetry at normal incidence, but instead arises from interference between two vertically hybridized leaky resonances.

A more quantitative comparison is obtained by analyzing the reflectivity spectra near the hybridization region. Figures~\ref{fig:exp}(e) and \ref{fig:exp}(f) show stacked experimental and simulated spectra, respectively. The quasi-BIC angle is highlighted in red, where one hybrid branch becomes strongly linewidth-suppressed while the other remains radiative. The resonance energies extracted from experiment and simulation are shown in Figures~\ref{fig:exp}(g) and \ref{fig:exp}(h), together with the analytical two-mode model. The model captures the avoided crossing and the hybrid-mode dispersions in both experiment and simulation. Figure~\ref{fig:exp}(i) further shows the calculated fraction of even parity for the two branches, confirming the continuous exchange of modal character across the avoided crossing.

The linewidth evolution in Figure~\ref{fig:exp}(j) provides direct evidence of Friedrich--Wintgen loss exchange. Because experimental linewidths can include additional broadening from finite illumination area, surface roughness, and in-plane scattering, we compare the analytical model to the radiative linewidths extracted from simulation. The lower branch exhibits a pronounced linewidth minimum near the quasi-BIC angle, while the upper branch remains comparatively broad. This behavior is well reproduced by the analytical two-mode model using a single set of parameters in the vicinity of the avoided crossing. Specifically, the fit is performed in the window $\theta_{AC} \pm 5^\circ$, assuming locally linear bare dispersions $\omega_m(\theta) = \omega_{AC} + \alpha_m(\theta - \theta_{AC})$ for $m = 1, 2$, and slowly varying bare radiative linewidths $\gamma_m$. The two independent coupling parameters are the effective coherent coupling $\kappa_{\mathrm{eff}}$ and the radiation overlap $\mu$. The measured off-$\Gamma$ quasi-BIC lies in the near-maximal-overlap regime, $\mu \simeq 1$; in this regime $\mathrm{Im}\Gamma \to 0$, so that $\kappa_{\mathrm{eff}}$ coincides with the physical coherent coupling $\kappa$ and the fit reduces to the real-coupling Friedrich--Wintgen form $\Gamma \simeq \sqrt{\gamma_1 \gamma_2}$.

\section{Case Study II: Bilayer Grating with Lateral Offset}

We now study the bilayer grating platform as a second case study of the same vertical-symmetry-breaking mechanism. Unlike the single-layer photonic crystal slab, where vertical symmetry is broken by superstrate/substrate-index contrast or partial etching, the bilayer grating realizes vertical symmetry breaking through a lateral displacement between two patterned layers. This laterally offset bilayer is the same platform previously used to demonstrate reconfigurable optical singularities and topological radiation-asymmetry control~\cite{Ni2024,Zhuang2024}; here we revisit it through the unified direction-resolved radiation-vector picture developed above, which places the symmetry-protected BIC, the (quasi-)UGRs, and the quasi-BICs on a common footing. The structure consists of two identical dielectric gratings embedded in a homogeneous environment with refractive index $n_{\mathrm{env}}=1.5$. The grating material has refractive index $n=2.8$, the period is $a=800~\mathrm{nm}$, and the height of each grating layer is $h=400~\mathrm{nm}$. The filling factor is defined as $FF=w/a$, where $w$ is the dielectric ridge width within one period; in the simulations, $FF=0.3$, corresponding to $w=240~\mathrm{nm}$. For the RCWA absorption spectra, a weak numerical probe loss was introduced by assigning an imaginary refractive-index component of $10^{-7}$ to the grating material. This artificial absorption channel is used only to reveal the resonant features in the absorption map and is sufficiently small that the modal dispersion remains essentially unchanged. The top and bottom gratings are laterally displaced by $d_x$, which modifies the near-field modal overlap and the relative amplitude and phase of the radiation emitted into the upper and lower half-spaces. Importantly, a nonzero displacement $d_x$ breaks both the vertical mirror symmetry ($z\to -z$) and the in-plane mirror symmetry ($x\to -x$), while preserving the inversion symmetry $(x,z)\to(-x,-z)$. The breaking of vertical mirror symmetry lifts the protection of the band crossing, whereas the residual inversion symmetry, combined with reciprocity, enforces a strict antisymmetry of the direction-resolved radiation in momentum space: the directional contrast obeys $\mathcal{D}_{\mathbf c}(-k_x)=-\mathcal{D}_{\mathbf c}(k_x)$, so that the preferred emission side reverses under momentum inversion $k_x\to -k_x$~\cite{Zhou2016}. Thus, the shifted bilayer implements the same two-mode non-Hermitian mechanism discussed above~\cite{Letartre2022}, with $d_x/a$ acting as the normalized symmetry-breaking control parameter.

\begin{figure*}[ht!]
    \centering
    \includegraphics[width=1\linewidth]{bilayer_grating.jpg}
    \caption{\textbf{From a symmetry-protected BIC to (quasi-)UGRs and quasi-BICs in a laterally shifted bilayer grating.}
    (a) Schematic of the bilayer grating. The lateral displacement $d_x$ between the upper and lower gratings breaks vertical symmetry.
    (b) Absorption spectrum for the aligned bilayer, $d_x/a=0$, where both the vertical ($z\to -z$) and in-plane ($x\to -x$) mirror symmetries are preserved. A symmetry-protected BIC, denoted s-BIC, appears at the $\Gamma$ point, protected by the in-plane $x\to -x$ mirror symmetry.
    (c) Absorption spectra for $d_x/a=0.05$, showing inequivalent responses for top and bottom incidence. Top incidence is labeled in red and bottom incidence in blue. $\mathrm{UGR}_1$ and $\mathrm{UGR}_2$ appear for top incidence, while $\mathrm{UGR}_1'$ and $\mathrm{UGR}_2'$ appear for bottom incidence, corresponding to nearly one-sided radiation cancellation for opposite incidence directions and opposite signs of $k_x$.
    (d) Absorption spectra for $d_x/a=0.124$, where q-BIC and q-BIC$^*$ appear due to strong suppression of the total radiative decay.
    (e) Representative $E_y$ field profiles of the s-BIC, q-BIC, q-BIC$^*$, $\mathrm{UGR}_1$, $\mathrm{UGR}_1'$, $\mathrm{UGR}_2$, and $\mathrm{UGR}_2'$, illustrating the evolution from a symmetry-protected BIC to (quasi-)UGRs and quasi-BICs. The black arrows indicate the dominant radiation direction of the (quasi-)UGR modes.
    (f) Quality-factor map, $\log_{10}(Q)$, in the $(k_x,d_x/a)$ parameter space. Yellow stars mark the high-$Q$ s-BIC, q-BIC, and q-BIC$^*$ states.
    (g) Directional-contrast map, $\mathcal{D}_{\mathbf c}$, in the same parameter space. The dashed horizontal line marks $d_x/a=0.05$, corresponding to the spectra in panel (c). Cyan diamonds mark the (quasi-)UGR states, which lie near extrema of $|\mathcal{D}_{\mathbf c}|$. Quasi-BICs and (quasi-)UGRs are related but distinct interference conditions: quasi-BICs suppress the total radiative decay, whereas (quasi-)UGRs suppress radiation predominantly into one half-space while maintaining finite radiation into the opposite half-space. A weak numerical probe loss, corresponding to an imaginary refractive-index component of $10^{-7}$ in the grating material, was included in the RCWA absorption calculations.}
    \label{fig:bilayer}
\end{figure*}

Figure~\ref{fig:bilayer} summarizes how lateral-shift-induced vertical symmetry breaking controls direction-resolved radiation, quasi-UGR formation, and quasi-BIC formation in the bilayer grating. Figure~\ref{fig:bilayer}(a) shows the bilayer geometry and defines the lateral displacement $d_x$. The two leaky bands considered here are both TE-polarized: they are the even and odd combinations, with respect to the vertical mirror $z\to -z$, of the fundamental TE guided resonances of the two identical grating layers, and thus play the role of the two opposite-vertical-parity modes of the general framework. In the aligned configuration ($d_x/a=0$), the structure preserves both the vertical ($z\to -z$) and in-plane ($x\to -x$) mirror symmetries, and the top- and bottom-incident spectra are equivalent. This symmetric reference is characterized by two key features [Figure~\ref{fig:bilayer}(b)]. First, the two bands have opposite vertical parity, so the vertical mirror symmetry forbids their mutual coupling and their dispersions simply cross. Second, a symmetry-protected BIC, denoted s-BIC, appears at the $\Gamma$ point: it is protected by the in-plane $x\to -x$ mirror symmetry, since at normal incidence the corresponding band is odd under $x\to -x$ while the zeroth-order radiation channel is even, so its coupling to the continuum is symmetry-forbidden. The band crossing and the s-BIC thus serve as the two reference features before the lateral shift is introduced.

Introducing a finite lateral shift breaks both mirror symmetries simultaneously, while preserving inversion symmetry~\cite{Zeng2021}, and each broken symmetry acts on one of these reference features. Breaking vertical mirror symmetry lifts the protection of the band crossing and converts it into an avoided crossing; this continuous evolution corresponds to a weak-to-strong coupling transition, with an EP appearing at the transition. The corresponding eigenfrequency evolution and complex-gap topology are presented in the Supplementary Material. Breaking the in-plane mirror symmetry, in turn, removes the symmetry protection of the s-BIC, allowing it to acquire a finite linewidth and to evolve into the quasi-UGRs and quasi-BICs discussed below.

When a finite lateral shift $d_x/a=0.05$ is applied, the top and bottom radiation channels become inequivalent. Figure~\ref{fig:bilayer}(c) shows that the spectra for top and bottom incidence are no longer identical, and strongly directional (quasi-)UGRs appear at off-$\Gamma$ momenta. The top-incident spectrum contains $\mathrm{UGR}_1$ and $\mathrm{UGR}_2$, while the bottom-incident spectrum contains the corresponding partners $\mathrm{UGR}_1'$ and $\mathrm{UGR}_2'$. These states remain leaky, but radiation is strongly, and in the ideal UGR limit completely, suppressed into one half-space while remaining finite into the opposite half-space. The four labeled states represent near side-resolved radiation-cancellation conditions for opposite incidence directions and opposite signs of $k_x$, showing that the preferred radiation direction reverses under momentum reversal and incidence reversal.

Increasing the shift to $d_x/a=0.124$ produces a different interference condition. As shown in Figure~\ref{fig:bilayer}(d), q-BIC and q-BIC$^*$ appear at opposite finite momenta. These quasi-BICs emerge in the vicinity of the avoided crossing between the two hybridized bands, as expected from the general behavior of non-Hermitian hybridization. In contrast to (quasi-)UGRs, where only one side-resolved radiation channel is suppressed, these quasi-BICs correspond to strong suppression of the total radiative decay. In the direction-resolved picture, the q-BICs can be interpreted as the near-coalescence of opposite quasi-UGRs conditions, where radiation into the upper and lower half-spaces is suppressed simultaneously.

The representative $E_y$ field profiles in Figure~\ref{fig:bilayer}(e) illustrate this evolution from the s-BIC state, to the strongly directional (quasi-)UGR states, and finally to the q-BIC and q-BIC$^*$ states. The black arrows indicate the dominant radiation direction of the (quasi-)UGR modes, confirming their strongly one-sided radiation character. In particular, $\mathrm{UGR}_1$ and $\mathrm{UGR}_2'$ radiate preferentially upward, whereas $\mathrm{UGR}_1'$ and $\mathrm{UGR}_2$ radiate preferentially downward. These field profiles show that the lateral shift not only changes the radiative coupling strength, but also reshapes the hybrid modal profiles that control the relative phase and amplitude of the upward and downward radiation channels.

Figures~\ref{fig:bilayer}(f) and \ref{fig:bilayer}(g) provide a global view in the $(k_x,d_x/a)$ parameter space. The quality-factor map in Figure~\ref{fig:bilayer}(f) tracks the high-$Q$ states marked by yellow stars: the symmetry-protected s-BIC at $k_x=0$ on the $d_x/a=0$ axis, and the finite-momentum q-BIC and q-BIC$^*$ that emerge at larger displacement. The directional-contrast map, $\mathcal{D}_{\mathbf c}$, in Figure~\ref{fig:bilayer}(g) highlights the (quasi-)UGR conditions. As imposed by the residual inversion symmetry, this contrast is antisymmetric in momentum, $\mathcal{D}_{\mathbf c}(-k_x)=-\mathcal{D}_{\mathbf c}(k_x)$, so the map is odd about $k_x=0$. The dashed horizontal line marks $d_x/a=0.05$, corresponding to the spectra in Figure~\ref{fig:bilayer}(c), and the cyan diamonds locate $\mathrm{UGR}_1$, $\mathrm{UGR}_2$, $\mathrm{UGR}_1'$, and $\mathrm{UGR}_2'$ near extrema of $|\mathcal{D}_{\mathbf c}|$. Crucially, these quasi-UGR points sit near the extrema of the directional-contrast map but away from the high-$Q$ ridges, whereas the q-BIC markers sit on the high-$Q$ ridges at $\mathcal{D}_{\mathbf c}\simeq0$. This separation between the two sets of markers makes explicit that quasi-BICs and quasi-UGRs impose different conditions on the same hybrid radiation vector: a quasi-BIC minimizes the total radiative decay ($\gamma_{\mathbf c}^{t}+\gamma_{\mathbf c}^{b}\to0$), whereas a (quasi-)UGR maximizes the directional contrast ($|\mathcal{D}_{\mathbf c}|\to1$) by cancelling, or nearly cancelling, radiation into a single half-space.

We emphasize an important distinction from the partially etched single-layer slab of Sec.~\ref{sec:numerics}. There, the quasi-BIC forms a robust line in parameter space, persisting as an extended high-$Q$ ridge over a broad range of etching ratios [Figure~\ref{fig:simul_contrast}(j)] because a near-collinear radiation-vector overlap is maintained over a finite parameter window. Here, by contrast, both the symmetry-protected BIC and the quasi-BICs are localized points in the $(k_x,d_x/a)$ parameter space: the s-BIC at $(k_x,d_x/a)=(0,0)$ and the q-BIC/q-BIC$^*$ at isolated finite displacements, in the same fashion as the symmetry-protected BIC is localized at the origin. This localization is topological in origin: each BIC or quasi-BIC can be viewed as the merging or near-merging point of four (quasi-)UGR branches---two radiating predominantly upward and two radiating predominantly downward---so that the surrounding directional-contrast map $\mathcal{D}_{\mathbf c}(k_x,d_x/a)$ develops a four-quadrant $\pm$ texture [Figure~\ref{fig:bilayer}(g)]. The BIC and quasi-BIC thus appear as the central singularity of this quadrant texture, i.e., a singularity of the radiation-asymmetry pseudo-polarization defined by the degree of asymmetric radiation~\cite{Zhuang2024,Yuan2026}. This pointlike, singularity-pinned character is fundamentally distinct from the extended quasi-BIC line of the robust partial-etching configuration.

These results reveal a continuous radiation-interference pathway from a symmetry-protected BIC to quasi-UGRs and quasi-BICs. The aligned bilayer supports the s-BIC because the in-plane $x\to -x$ mirror symmetry forbids coupling of the corresponding mode to the radiation continuum at $\Gamma$. Once a lateral shift is introduced, both the in-plane and vertical mirror symmetries are broken while inversion is preserved, the symmetry protection is lifted, and the radiation vector separates into side-resolved cancellation conditions, producing quasi-UGRs where one half-space is dark or nearly dark. At larger displacement, opposite directional-cancellation conditions nearly coalesce, leading to q-BIC and q-BIC$^*$ formation through strong suppression of the total radiative decay. Therefore, quasi-UGRs and quasi-BICs are not independent effects, but two limits of the same vertical-symmetry-breaking-induced radiation-vector interference.

Finally, it is worth commenting on the robustness of the quasi-UGRs observed in both case studies. In the partially etched single-layer slab and in the laterally offset bilayer alike, high directional contrast is not a fine-tuned accident: as one vertical-symmetry-breaking parameter is varied---the etching ratio in Sec.~\ref{sec:numerics} or the lateral shift $d_x$ here---a one-sided or nearly one-sided radiation condition can always be recovered by simultaneously adapting a second parameter, here the in-plane wavevector $k_x$. This behavior is the direct manifestation of the codimension-one nature of UGRs established in Sec.~\ref{sec:UGR}: because one-sided or nearly one-sided radiation cancellation is controlled by a single real condition for one dominant open channel per side, a generic perturbation does not destroy a quasi-UGR but merely displaces it, so that quasi-UGRs persist as continuous ridges in parameter space. Equivalently, starting from a bound state in the continuum---here the symmetry-protected s-BIC---a continuous family of quasi-UGRs can be generated by tuning a single parameter, in agreement with the rigorous parametric analysis of unidirectional guided resonances in periodic structures~\cite{YuanLu2025}. This parametric robustness explains why both vertical-symmetry-breaking platforms reliably yield strongly directional emission over extended regions of their design space, even though the underlying BICs and quasi-BICs remain localized in parameter space.

\section{Conclusion}
We have developed a unified two-mode non-Hermitian framework for leaky Bloch resonances in photonic crystal slabs with broken vertical mirror symmetry. The central result is that vertical-symmetry breaking simultaneously modifies near-field modal hybridization and radiative-channel interference between resonances that originate from opposite vertical parity sectors. Within one minimal model, this produces a continuous weak-to-strong coupling transition, with EPs appearing at the boundary between frequency crossings and avoided crossings, while the same hybridization enables Friedrich-Wintgen redistribution of radiative loss.

The framework separates the spectral and radiative content of the problem. The effective spectral coupling determines the weak-to-strong transition and the EP condition, whereas the effective loss-exchange parameter determines the maximum linewidth redistribution accessible through radiative interference. A deeply suppressed hybrid linewidth therefore does not require a large avoided-crossing gap, and quasi-BIC behaviour can occur in the weak-coupling regime, at an EP, or in the strong-coupling regime. In the ideal limit, complete cancellation of the hybrid radiation vector produces a FW BIC; more generally, incomplete but efficient cancellation produces a high-$Q$ quasi-BIC over a finite parameter range.

Resolving the same radiation vector into its top and bottom components provides a complementary direction-resolved description. A quasi-BIC corresponds to suppression of the total radiative decay into all open channels, whereas a UGR or quasi-UGR corresponds to cancellation, or near-cancellation, of radiation into only one half-space. EPs, quasi-BICs, and directional guided resonances are therefore distinct manifestations of a common hybrid radiation problem: the EP is a spectral singularity, the quasi-BIC is a total-radiation interference condition, and the UGR is a side-resolved interference condition.

We validated this picture using a fabrication-compatible square-lattice SiN$_x$-on-SiO$_2$ PhC slab. Full-wave simulations show a continuous evolution from a frequency crossing in the vertically symmetric reference structure to an avoided crossing after vertical symmetry is broken through superstrate/substrate asymmetry. By tuning the superstrate index, the weak-to-strong coupling transition passes through an EP characterized by coalescence of the complex eigenfrequencies and the associated half-integer phase winding. Partial etching provides an additional degree of freedom for radiative-channel engineering and produces a broad high-$Q$ quasi-BIC regime with strong FW linewidth redistribution. Angle-resolved reflectivity measurements reveal an off-$\Gamma$ symmetry-breaking quasi-BIC near the hybridization region; its dispersion and linewidth evolution are reproduced by the same two-mode model, supporting its interpretation as a FW loss-exchange effect.

The direction-resolved simulations further show that quasi-BIC formation and strongly asymmetric radiation need not coincide. In the partially etched single-layer slab, a quasi-BIC minimizes the total radiative loss, whereas a quasi-UGR maximizes top-bottom emission contrast while retaining radiation into the opposite half-space. In the laterally shifted bilayer grating, where vertical and in-plane mirror symmetries are broken simultaneously, the same side-resolved radiation-vector description connects the symmetry-protected BIC, quasi-UGRs, and quasi-BICs. The quasi-UGRs form continuous high-contrast ridges in parameter space, consistent with their codimension-one character, while the BIC and quasi-BIC states remain localized near the points where the relevant directional-cancellation conditions coincide.

More broadly, the framework applies to a wide class of vertically asymmetric photonic structures, including asymmetric cladding configurations, partially etched slabs, tilted-sidewall gratings, and multilayer or hetero-bilayer platforms. It provides a practical basis for distinguishing and independently engineering spectral hybridization, total-radiation suppression, and one-sided radiation cancellation. Our results open clear perspectives for designing robust quasi-BIC windows via radiative-channel engineering, leveraging the topological features of EPs in parameter space, and extending vertical-symmetry-controlled coupling to active and reconfigurable devices for directional emission~\cite{Ha2018,MermetLyaudoz2023}, low-threshold lasing~\cite{Gu2022,Ferrier2022}, and non-Hermitian photonic functionality.

\textit{Acknowledgments.}---The authors would like to thank the staff from the NanoLyon Technical Platform for helping and supporting in all nanofabrication processes, the Consortium Lyon Saint-Etienne de Microscopie (CLYM, FED 4092) for the access to the microscopes and David Albertini for having done the AFM image of the structure. This work is supported by the French National Research Agency (ANR) under the projects POPEYE and EMIPERO, and by Vietnam National Foundation for Science and Technology Development (NAFOSTED) under grant number 103.03-2024.16. H.S.N. acknowledges financial support from the French National Research Agency (ANR) under the projects POLAROID (ANR-24-CE24-7616-01), SUPER-HERO (ANR-25-CE24-4066), STRONG-NANO (ANR-22-CE24-0025), and CHOCOLAT (ANR-22-CE24-0025).

\bibliography{Ref}
\include{supp2}
\end{document}

%% file: supp2.tex
\onecolumngrid

\begin{center}
		\textbf{\large --- Supplementary Material ---\\ Unified Weak-to-Strong Coupling Transitions and Radiation Interference Induced by Vertical-Symmetry Breaking in Photonic Crystal Slabs}\\
		\medskip
	\end{center}

\setcounter{equation}{0}
\setcounter{figure}{0}
\setcounter{table}{0}
\setcounter{page}{1}
\setcounter{section}{0}

\renewcommand{\theequation}{S\arabic{equation}}
\renewcommand{\thefigure}{S\arabic{figure}}
\renewcommand{\bibnumfmt}[1]{[S#1]}
\renewcommand{\vec}[1]{\boldsymbol{#1}}
\renewcommand{\correct}[2]{{\color{red}\sout{#1}\,}{\color{blue}#2}}
\vspace{3cm}

\section{Far-field polarization texture near the symmetry-breaking quasi-BIC}

A genuine photonic bound state in the continuum is commonly associated with a zero of the radiative amplitude in momentum space. At such a point, the radiative quality factor diverges, $Q_{\mathrm{rad}}\rightarrow\infty$, and the far-field polarization becomes undefined, giving rise to a polarization singularity with a vortex-like winding of the polarization vector. In contrast, the state discussed in the main text is the off-$\Gamma$ symmetry-breaking quasi-BIC observed in the partially etched single-layer PhC slab. This quasi-BIC is generated by Friedrich--Wintgen radiative loss exchange between two hybridized leaky resonances of opposite vertical parity. Its radiative linewidth is strongly suppressed but remains finite because the destructive interference between the two radiation channels is not perfectly complete. Therefore, this state should be distinguished from the symmetry-protected $\Gamma$-point BICs, whose radiation is forbidden by symmetry, and from an ideal FW BIC, which would require perfect radiative cancellation.

\begin{figure}[h!]
\begin{center}
\includegraphics[width=14cm]{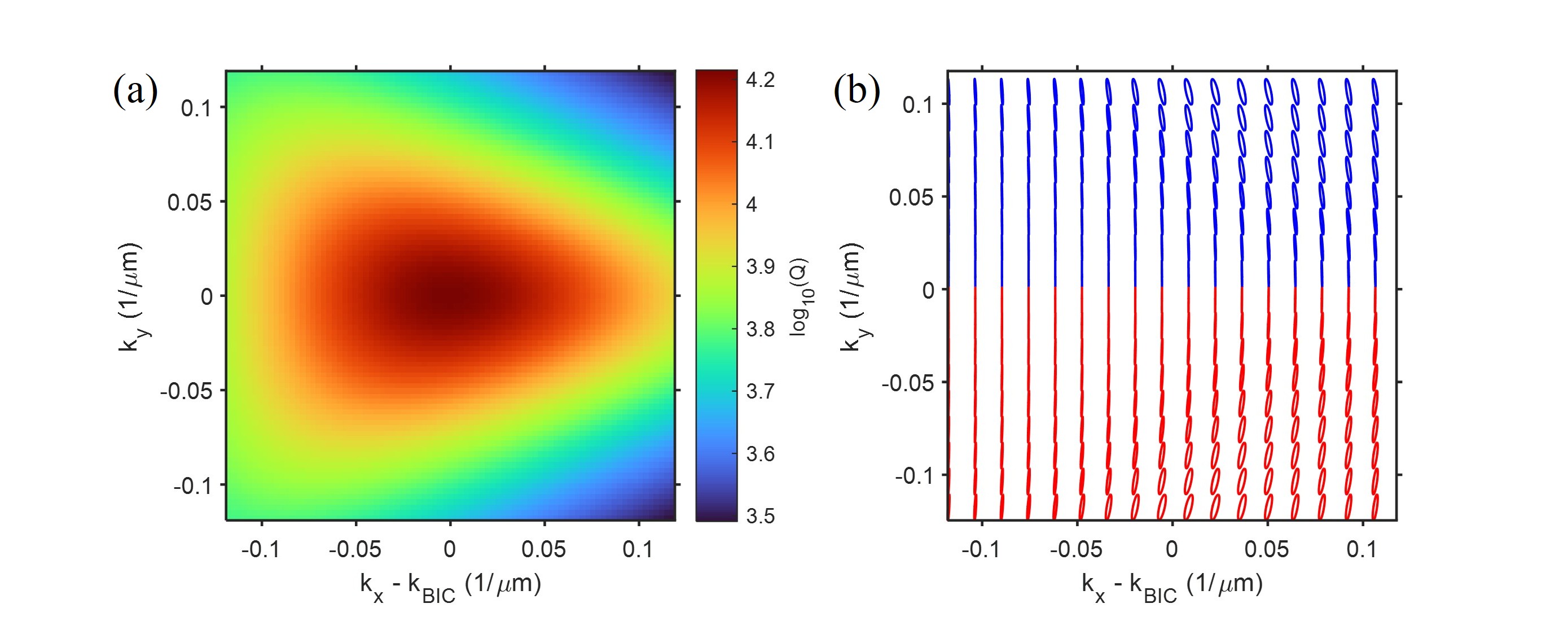}
\end{center}
\caption{\label{fig5}{(a) Calculated radiative quality factor $Q_{\mathrm{rad}}$ in the vicinity of the quasi-BIC point in the $(k_x,k_y)$ plane using FEM. The quality factor is strongly enhanced but remains finite, confirming that the state is a quasi-BIC rather than a genuine BIC. (b) Corresponding far-field polarization ellipses. The polarization texture evolves smoothly and does not form a vortex, in contrast to the polarization singularity expected for a true BIC with $Q_{\mathrm{rad}}\rightarrow\infty$.
}}
\label{fig:sm_qbic_pol_mapping}
\end{figure}

To verify this distinction, we calculate the far-field radiation properties in the vicinity of the quasi-BIC point. Figure~\ref{fig:sm_qbic_pol_mapping}(a) shows the radiative quality factor $Q_{\mathrm{rad}}$ in the $(k_x,k_y)$ plane. A pronounced maximum is observed near the quasi-BIC, but $Q_{\mathrm{rad}}$ remains finite rather than diverging, confirming that the state is not an exact BIC. This behavior is further confirmed by the polarization ellipses in Figure~\ref{fig:sm_qbic_pol_mapping}(b), which evolve smoothly in the surrounding momentum space and do not form a vortex. These results demonstrate that the off-$\Gamma$ linewidth suppression discussed in the main text corresponds to a high-$Q$ quasi-BIC generated by incomplete Friedrich--Wintgen destructive interference, rather than to a genuine far-field polarization singularity.

These results show that the off-$\Gamma$ linewidth suppression observed in the main text should be interpreted as a high-$Q$ quasi-BIC rather than a topological far-field polarization singularity. The quasi-BIC originates from strong but incomplete Friedrich--Wintgen destructive interference between hybridized modes of opposite vertical parity.

\section{Symmetry-breaking quasi-BIC in p-polarization ($H_y$ field)}

\begin{figure}[ht!]
\begin{center}
\includegraphics[width=14cm]{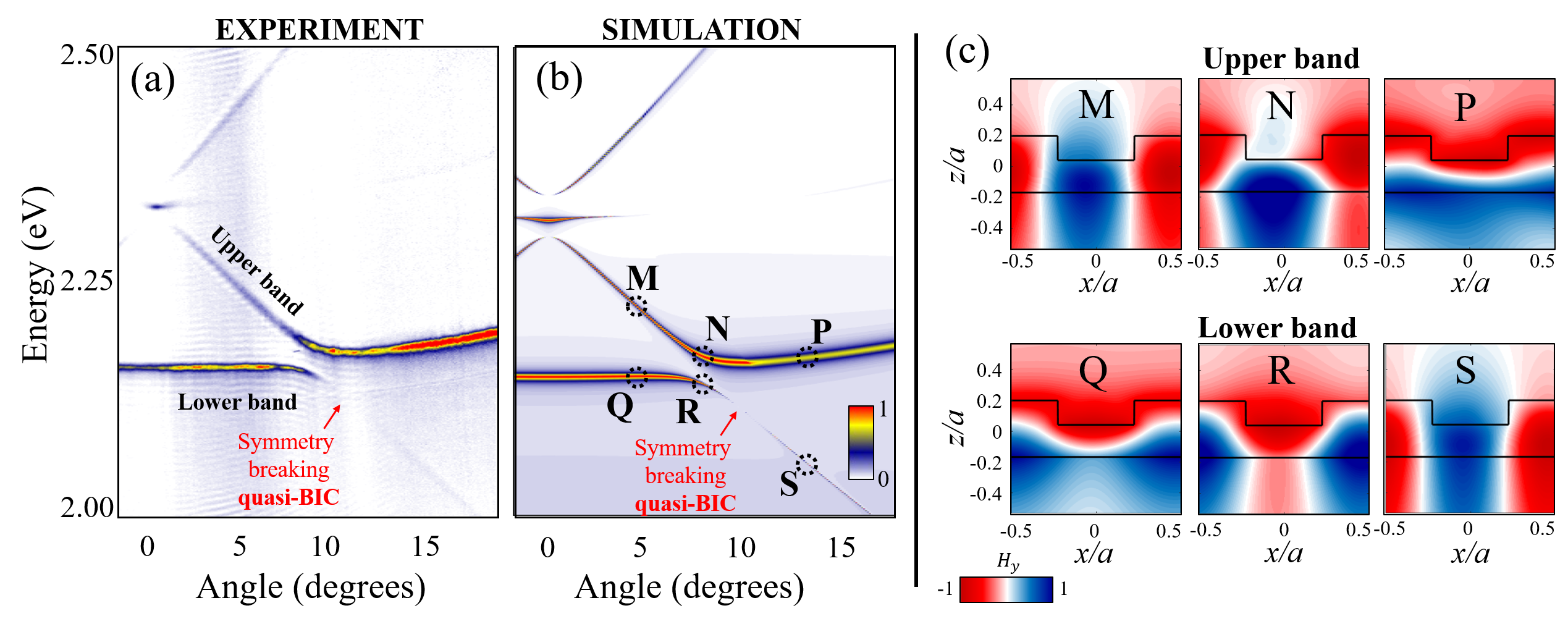}
\end{center}
\caption{\label{fig7}{(a) Experimental measurement and (b) Numerical simulations of the p-polarized angle-resolved reflectivity. (c) Distribution of the magnetic field component $H_y(x,y=0,z)$ on different points of the upper and lower modes of (b).}}
\label{fig:SP}
\end{figure}

We present the experimental angle-resolved reflectivity measured in $p$ polarization, corresponding to the $H_y$ field, together with the corresponding numerical simulation [Figures~\ref{fig:SP}(a) and \ref{fig:SP}(b), respectively]. The measurements are performed on the same partially etched PhC slab sample discussed in the main text, where the $p$-polarized response is used to further confirm the symmetry-breaking quasi-BIC mechanism. Similar to the quasi-BIC observed in $s$ polarization, the $p$-polarized spectra reveal the same hybridization physics. First, far from the avoided crossing point, the two modes recover predominantly even-like character at points M and S, or odd-like character at points Q and P. Second, near the avoided crossing point, the modes exhibit balanced mixing between the odd- and even-like components, as shown at points N and R. Third, a symmetry-breaking quasi-BIC, identified by the vanishing of the Fano resonance, is observed in the vicinity of the avoided crossing point on the lower branch. These results show that the experimentally observed quasi-BIC is not restricted to the $s$-polarized response, but also appears in the $p$-polarized measurement of the same structure.

We note that in the case of p-polarized, the linewidths of uncoupled modes are very different: $\gamma_{e}\gg\gamma_{o}$ [see Figure~\ref{fig:SP}(a)], thus the angular position $\theta_{BIC}$ of the symmetry-breaking quasi-BIC is greatly shifted with respect to $\theta_{AC}$ of the avoided crossing point with $|\theta_{AC}-\theta_{BIC}|=2^\circ$. This is very different from the case of s-polarized in which $\gamma_{e}\approx\gamma_{o}$, leading to $|\theta_{AC}-\theta_{BIC}|=0.5^\circ$.

\section{Exceptional Point at the Weak-to-Strong Coupling Transition in a Laterally Offset Bilayer Grating.}

\begin{figure}[ht!]
    \centering
    \includegraphics[width=1\linewidth]{bilayer_ep.jpg}
   \caption{
Numerical demonstration of the weak-to-strong coupling transition in the bilayer grating, with an exceptional point (EP) emerging at the boundary. Panels (a)--(c) show the real and imaginary parts of the complex eigenfrequencies calculated for different lateral shifts $d_x/a$ between the top and bottom gratings. The three panels illustrate the continuous transition from the weak-coupling regime to the strong-coupling regime through an EP: in (a) at $d_x/a = 0.0225$, the real parts cross while the imaginary parts remain separated; in (b) at $d_x/a = 0.0225368$, both the real and imaginary parts coalesce at the EP; and in (c) at $d_x/a = 0.0226$, the real parts undergo an avoided crossing while the imaginary parts exchange between the two branches. Panels (d) and (e) show, respectively, the magnitude and phase of the complex eigenfrequency gap when both the in-plane wavevector $k_x$ and the lateral shift $d_x/a$ are varied. The EP is identified by the vanishing complex gap in panel (d) and by a $\pm\pi$ phase winding in panel (e), corresponding to a half-integer topological charge of $\pm 1/2$. Geometrical and material parameters are given in the main text.
}
    \label{fig:bilayer_ep}
\end{figure}

The universal picture developed above predicts that the transition from weak to strong coupling occurs through an exceptional point, where both the complex eigenfrequencies and eigenvectors coalesce. In Case Study I, this transition was demonstrated in a single-layer PhC slab by tuning the superstrate refractive index. Here, in Case Study II, we show that the same weak-to-strong coupling scenario also appears in the laterally shifted bilayer grating, with the normalized lateral displacement $d_x/a$ acting as the symmetry-breaking control parameter.

Figure~\ref{fig:bilayer_ep} confirms this transition. The complex eigenfrequencies evolve from the weak-coupling regime, where the real parts cross while the imaginary parts remain separated [Figure~\ref{fig:bilayer_ep}(a), $d_x/a=0.0225$], to the transition point, where both real and imaginary parts coalesce at an exceptional point [Figure~\ref{fig:bilayer_ep}(b), $d_x/a=0.0225368$], and finally to the strong-coupling regime, where the real parts anticross while the imaginary parts exchange between the two branches [Figure~\ref{fig:bilayer_ep}(c), $d_x/a=0.0226$]. The EP is further verified by the complex-gap maps: the magnitude of the eigenfrequency splitting vanishes at an isolated point in the $(k_x,d_x/a)$ parameter space [Figure~\ref{fig:bilayer_ep}(d)], while the phase winds by $\pm\pi$ around this point [Figure~\ref{fig:bilayer_ep}(e)], confirming the associated half-integer topological charge ±1/2.